\documentclass{aa}

\usepackage{graphicx}

\usepackage{placeins} 

\usepackage{txfonts}
\usepackage{xcolor}
\usepackage{hyperref}
\hypersetup{
colorlinks=true,
linkcolor=blue,
filecolor=blue,
urlcolor=blue,
citecolor=blue}

\begin{document}

   \title{Confirmation of a ring structure in the disk around MP  Mus (PDS 66) with ALMA Band 7 observations}

   \author{Aurora Aguayo
          \inst{1}
          \and
          Claudio Caceres\inst{2}
          \and
          Zhen Guo\inst{1,3}
          \and
          Matthias R. Schreiber\inst{4}
          \and
          Álvaro Ribas\inst{5}
          \and
          Joel H. Kastner\inst{6,7,8}
          \and
          Lucas A. Cieza\inst{9,10}
          \and
          Sebastián Pérez\inst{10,11,12}
          \and
          Héctor Cánovas\inst{13}
          \and
          Daniela Rojas Bozza\inst{2}
          \and
          D. Annie Dickson-Vandervelde\inst{7,8}
          \and
          William Grimble\inst{14,15}
          \and
          Alejandro Santamar\'ia-Miranda\inst{16}
          }

\institute{Instituto de Física y Astronomía, Facultad de Ciencias, Universidad de Valparaíso, Av. Gran Bretaña 1111, Valparaíso, Chile
              \email{aurora.aguayov@gmail.com, zhen.guo@uv.cl}
         \and
             Instituto de Astrofisica, Departamento de Fisica y Astronomia, Facultad de Ciencias Exactas, Universidad Andres Bello. Av. Fernandez Concha 700, Las Condes, Chile
        \and
             Millennium Institute of Astrophysics, Nuncio Monse{\~n}or Sotero Sanz 100, Of. 104, Providencia, Santiago, Chile
        \and
             Departamento de F\'isica, Universidad T\'ecnica Federico Santa Mar\'ia, Avenida Espa\~na 1680, Valpara\'iso, Chile
        \and
             Institute of Astronomy, University of Cambridge, Madingley Road, Cambridge, CB3 0HA, UK
        \and
             Center for Imaging Science, Rochester Institute of Technology, Rochester, NY 14623, USA; jhk@cis.rit.edu
        \and
             School of Physics and Astronomy, Rochester Institute of Technology, Rochester, NY 14623, USA
        \and
             Laboratory for Multiwavelength Astrophysics, Rochester Institute of Technology
        \and
             Instituto de Estudios Astrofísicos, Facultad de Ingeniería y Ciencias, Universidad Diego Portales,  Av. Ejercito 441, Santiago, Chile
        \and
             Millennium Nucleus on Young Exoplanets and their Moons (YEMS), Chile
        \and
             Departamento de Física, Universidad de Santiago de Chile, Av. Víctor Jara 3493, Santiago, Chile
        \and
             Center for Interdisciplinary Research in Astrophysics $\&$ Space Exploration (CIRAS), Universidad de Santiago de Chile, Chile
        \and
            Telespazio UK S.L. for European Space Agency (ESA), Camino bajo del Castillo, s/n, Urbanización Villafranca del Castillo, Villanueva de la Cañada, 28692 Madrid, Spain
        \and
             Department of Physics and Astronomy, Vanderbilt University, Nashville, TN 37235, USA
        \and
             First Center for Autism and Innovation, Vanderbilt University, 2414 Highland Avenue, Suite 115, Nashville, TN 37212, USA
        \and
             Departamento de Astronom\'ia, Universidad de Chile, Camino El Observatorio 1515, Las Condes, Santiago, Chile
             }

   \date{Received XXX; accepted XXX}

  \abstract
   {Young stellar objects (YSOs) are surrounded by protoplanetary disks, which are the birthplace of young planets. Ring and gap structures are observed among evolved protoplanetary disks, often interpreted as a consequence of planet formation.  }
   {The pre-Main Sequence (pre-MS) star MP Mus hosts one of the few known examples of protoplanetary disks within ~100 pc. Previously, a disk ring structure, with a radius of 80-85 au, was detected in scattered light via near-infrared coronographic/polarimetric imaging. This ring structure may be indicative of the disk clearing process. Although such ring structures were not seen in the ALMA Band 6 images, some features were detected at $\sim$50 au.}
   {In this paper, we analyzed new ALMA Band 7 observations of MP Mus in order to investigate the details of its disk substructures. }
   {By subtracting the continuum profile generated from Band 7 data, we discovered a ring structure in the Band 7 dust continuum image at $\sim$50 au. We calculated the overall dust mass as $28.4\pm2.8 M_{\oplus}$ at 0.89 mm and $26.3\pm2.6 M_{\oplus}$ at 1.3 mm and the millimeter spectral index $\alpha_{0.89-1.3mm} \sim 2.2 \pm 0.3$ between 0.89 mm and 1.3 mm. Moreover, we display the spatial distribution of the spectral index ($\alpha_{mm}$), estimating values ranging from 1.3 at the inner disk to 4.0 at a large radius. Additionally, we observed an extended gas disk up to $\sim$120 au, in contrast with a compact continuum millimeter extent of $\sim$60 au.}
   {We conclude that there are strong indicators for an active radial drift process within the disk. However, we cannot discard the possibility of a dust evolution process and a grain growth process as responsible for the outer disk structures observed in the ALMA continuum imaging.}

   \maketitle

\section{Introduction}
\label{sec:1}

Stars are born from molecular clouds. Due to the conservation of angular momentum, newborn stars are surrounded by dust and gas-rich circumstellar disks, known as protoplanetary disks, which are the formation sites of young planets. In the protoplanetary disk phase (gas-rich), dust particles grow and may form objects large enough to accrete gas from the disk in a runaway process. Also, substructures such as gaps and rings are generated, potentially by planet formation and/or other processes. After roughly ten million years, photoevaporation dissipates the gas disk, leaving behind a debris disk \citep[e.g.,][]{Williams&Cieza2011} which may then evolve into a mature planetary system.

The structure and evolution of the protoplanetary disk provide crucial information on planet formation \citep{Williams&Cieza2011, Andrews2020, Pinte2023}. High spatial resolution observations have discovered substructures inside the circumstellar disk, potentially indicating their ongoing planet formation process \citep[see][]{Andrews2018, Long2018}, as early as during the protostellar stage \citep{SeguraCox2020}. The bright and dark ring structures have been a topic of great debate as they are expected to be detected as a consequence of already formed (proto)planets on the disk \citep{Zhu2014, Perez2015}. However, there are other mechanisms that can also form or show these structures, without requiring the presence of protoplanets, such as an effect of dust opacity due to grain growth \citep{Birnstiel2015}, hydrodynamic instabilities \citep{Ward2000, TakahashiInutsuka2014}, magnetohydrodynamic instabilities \citep{Gressel2015, Flock2015}, fast pebble growth near condensation fronts \citep{Zhang2015}, and dust rings induced by sintering \citep{Okuzumi2016, OkuzumiTazaki2019, Hu2019}. Regardless of their origin, these ring structures are excellent places to accumulate and concentrate large dust grains \citep[e.g.,][]{Zhu2014, Flock2015, Ruge2016, Sierra2017, Sierra2019}, and their presence among disks of all ages \citep[e.g.,][]{VanderMarel2016, Andrews2018, Shi2024} becomes a great indicator that the formation of rings is a necessary first step for the formation of terrestrial planets \citep{Carrasco-Gonzalez2016, Kuwahara2024}.

The sizes of even the brightest and largest protoplanetary disks in any given molecular cloud are $\sim$100 au \citep{Williams&Cieza2011}, whilst the median size of most disks is $\sim$14 au \citep{Dasgupta2025}. For spatially resolving disks in nearby star formation regions ($\sim$140 pc), observations with subarcsecond resolutions are required. Thus, observations with the Atacama Large Millimeter/submillimeter Array (ALMA) became crucial to resolve structures in disks. In the past decade, ALMA has provided data that revolutionized our understanding of the spatial properties of protoplanetary disks. This includes constraints on the distribution of millimeter sized dust grains, the kinematic parameters of molecular gas, and the disk chemical structures \citep[see][]{Andrews2020, Huang2018, Oberg2021}. High resolution ALMA data will also assist us in distinguishing different disk clearing mechanisms (e.g., photoevaporation and grain growth), providing constraints to the theoretical models, and finally placing an accurate diagnosis of the evolutionary stage of the planetary system.

Here, we focus on submillimeter observations obtained with ALMA of the disk around MP Mus, also known as PDS 66 or Hen 3-892: A star of 1.2 $\pm$ 0.2 $M_{\odot}$, K1 spectral type and an age of 3.1 $\pm$ 0.9 Myr \citep{Asensio-Torres2021}. According to Gaia DR3, the distance of MP Mus is 97.8 pc \citep{Gaia2024}, placing it among the three known, actively accreting, near solar mass stars within 100 pc. The other two nearby, roughly solar-mass pre-main sequence (pre-MS) stars that display evidence for ongoing accretion are TW Hya and V4046 Sgr \citep{KastnerPrincipe2022}. The young nature of MP Mus was originally reported by \citet{Gregorio1992}, based on the infrared excess and a strong H$\alpha$ emission feature in the spectrum. This star is located in the nearest OB association Scorpius-Centaurus and was originally considered a member of the Lower Centaurus Cruz (LCC) subgroup \citep{Mamajek2002}, with an estimated age of 17 Myr. Later, MP Mus was classified as a member of the $\epsilon$ Chamaeleon Association (CA), which has a much younger age of 3-8 Myr \citep{Torres2008, Murphy2013, Dickson2021}. A younger age of the system, 7 Myr, is also supported by the strength of the Li absorption line at 6708 \AA$\,$ \citep{Weise2010} and the star’s position near the boundary of the $\epsilon$ CA and the youngest ($\sim$8 Myr) LCC subgroup \citep{Varga2024}. Applying contemporaneous near-UV observation from the Hubble Space Telescope (HST) and ground-based optical spectrum, \citep{Ingleby2013} measured the mass accretion rate as 1.3$\times 10^{-10} M_\odot$ yr${^{-1}}$, placing it at the lower mass accretion range of pre-MS stars.

The measurement of the circumstellar disk properties of MP Mus started almost two decades ago. First, the total dust mass of the disk has been estimated to be 5$\times 10^{-5} M_\odot$ or 1.6 $M_\oplus$ \citep{Carpenter2005}. Using the N-band (8 - 13 $\mu$m) spectra and infrared Spectral Energy Distribution (SED) up to 1 mm, \citet{Schutz2005a} discovered grain growth in the circumstellar disc. Later on, \citet{Bouwman2008} detected circumstellar silicate dust grains in the disk of MP Mus, applying mid-infrared spectra from the Spitzer satellite. The first scattered light image of MP Mus was taken by the NICMOS on HST \citep[][]{Cortes2009}, with an estimated disk radius of 170 au. \citet{Kastner2010} established the presence of a significant gaseous component within the MP Mus disk via detection of $^{12}$CO(3-2) emission and estimated the gas disk radius as $\sim$120 au. With the development of the laser guiding near-infrared observations, a disk ring was detected around 80 au by the Gemini Planet Imager \citep{Wolff2016}. The dip of the surface brightness between the bright optically thick inner disk and the ring structure at 80 au indicates an ongoing clearing process inside the disk plane, consistent with the classification of a transitional disk. Similarly, \citep{Avenhaus2018} discovered a ring structure at 85 au and an inclination of 30 deg using the SPHERE instrument located in ESO/VLT. In addition, no accreting sub-stellar component was detected by the H$\alpha$ emission inside the disk cavity down to the 12 mag contrast of the central star \citep{Zurlo2020}.

\cite{Ribas2023} presented ALMA observations of MP Mus down to 4 au resolution in the 1.3 mm bandpass (Band 6). They found an optically thick disk with a radius of 60$\pm$5 au, with the possible exception of a barely resolved outer ring at $\sim$50 au. The mismatch between the gap locations detected in scattered light and dust continuum indicates that the ring structure seen in the near-infrared might only be a shadow cast by a puffed inner disk. The molecular gas (e.g., $^{12}$CO) emission spreads widely to a radius of 130 au. They also measured the millimeter spectral index ($\alpha_{mm}$) and found an inhomogeneous distribution of the $\alpha_{mm}$ value. This result indicates that the optically thick emission arises from high albedo dust grains residing in the inner 30 au of the disk ($\alpha < 2$). Subsequently, \citet{Grimble2024} analyzed the structure and composition of MP Mus using the Spitzer IRS spectrum and ALMA Band 6 image, with radiative transfer models. In this work, the authors found that a density gap (i.e., a ring at 80 au) fits the observational SED better than disk shadowing, within the limitations of the modeling.

\section{Observations and data reduction}
\label{sec:2}

In this study, we present and analyze ALMA observations of MP Mus from two Cycle 5 programs at 1.3 mm (Band 6; 2017.1.01419.S P.I. Claudio Cáceres; 2017.1.01167.S P.I.: Sebastián Pérez) and one Cycle 8 program at 0.89 mm (Band 7; 2021.1.01205.S P.I.: Claudio Cáceres). For the continuum analysis, we focus on Band 7 observations, and for comparison, we used the extended configuration of 2017.1.01167.S in Band 6 since it has the best resolution between the two datasets. The rms of the compact configurations for both datasets are systematically higher than the one of the extended configurations. In addition, the compact configurations Band 7 dataset showed issues during the phase calibration. 
After combining the compact and extended configurations we did not see an improvement in the rms nor in the image quality, thus we decided to use only the extended configurations of Band 7 and Band 6 for the continuum analysis. However, for line analysis, we used the combined compact and extended configurations of each data set. The details of the different data sets are summarized in Tables \ref{tab:observation log} and \ref{tab:correlator config log}, including the corresponding correlator configurations.

For data reduction, we used the standard pipeline calibration provided by the ALMA staff using the Common Astronomy Software Applications package, \citep[CASA;][]{McMullin2007} version 5.6.1-8-10.14. We detected continuum emission from the disk with a high peak Signal to Noise ratio (S/N >\texttt{~100}). After this, we performed selfcalibration to the extended configurations for each band; first excluding all channels with emission lines and then, using \texttt{mtmfs} deconvolver, \texttt{Briggs} weighting, a \texttt{robust} value of 0.5 and -0.5, and \texttt{nterms}=2. In addition, selfcalibration was performed on the combined compact and extended configurations for the line analysis. To properly combine these datasets, we found the peak of the disk continuum emission by fitting a 2D Gaussian shape to each dataset. Then, we applied CASA tasks \texttt{fixvis} and \texttt{fixplanets} to set the phase center to a common coordinate to finally use \texttt{concat} task to concatenate both data sets. 

For Band 7, we performed a five-step phase only calibration. We sought solutions over 60~s intervals on the first four steps and solutions over 45~s on the fifth step. By doing so, we improved the root mean square (rms) noise by 60\% and S/N of the emission peak by 215\% (from 125 to 394) for a robust value of $r = 0.5$ and an improvement of 30\% on S/N (from 107 to 141) for a robust value of $r = -0.5$. For Band 6 observations, we performed a similar four step phase only calibration and sought solutions over 35~s for the first and second steps, and solutions over 25~s for the third and fourth steps.  In this case, the rms noise is reduced by 80\% and S/N is improved by 800\% (from $\sim$40 to $\sim$320) for both robust values: $r = 0.5$ and $r = -0.5$.

\begin{figure*}
\includegraphics[width=7in]{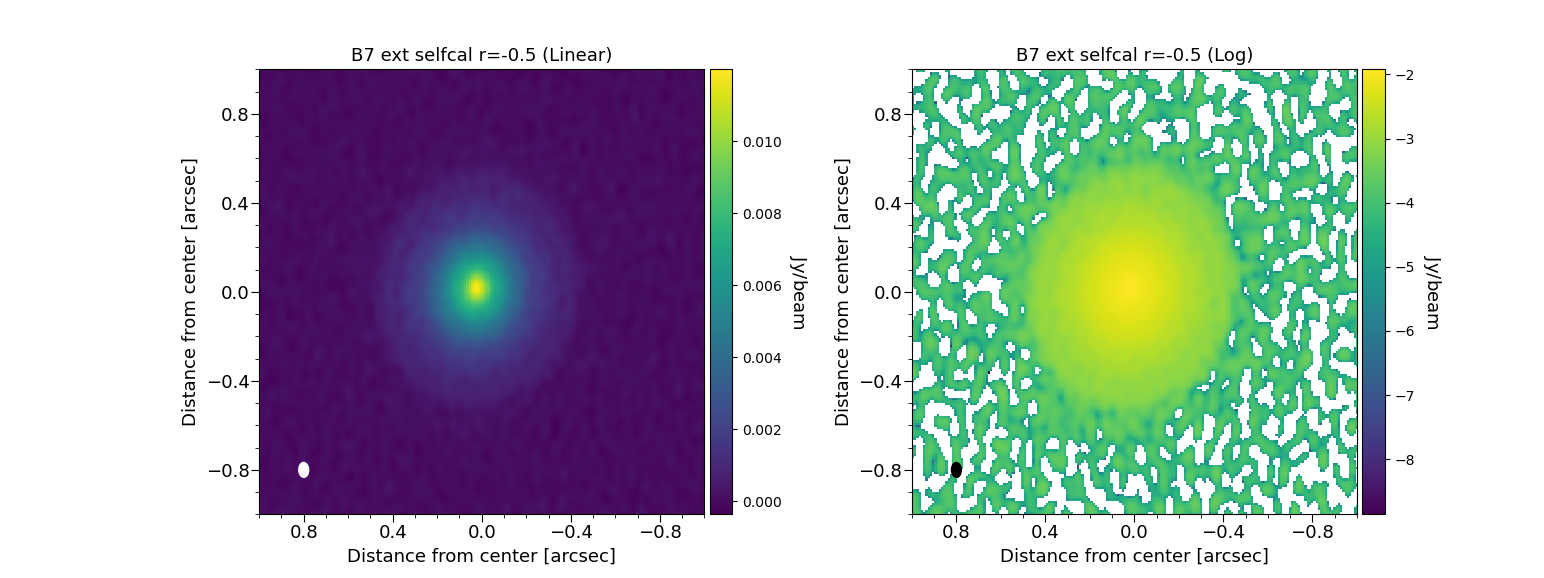}
\caption{ALMA Band 7 continuum images of MP Mus at 0.89 mm using the selfcalibrated extended configuration with a robust value of r = -0.5. The synthesized beam size (white ellipse) is shown in the bottom left corner. \textit{Left:} Linear scale; \textit{Right:} Logarihtmic scale.}
\label{fig:Continuum B7 ext selfcal r-05}
\end{figure*}

\section{Results}
\label{sec:3}

The ALMA Band 7 observation provides a high spatial resolution of 0.07$\times$0.05" that complements the previous Band 6 observations that reach a resolution of 0.06$\times$0.04" \citep[][]{Ribas2023}, both with a robust value of -0.5. In this section, we will analyze the disk structures from these Band 6 and 7 observations, in terms of the continuum images and radial profiles, and their CO spectral line counterparts.

\subsection{Dust continuum}
\label{sec:3.1}

\subsubsection{Continuum images}
\label{sec:3.1.1}

We present synthesized continuum images at 0.89 mm (Band 7) and 1.3 mm (Band 6) from the selfcalibrated data described in Sect. \ref{sec:2}. We applied the \texttt{tclean} algorithm with \texttt{ mtmfs} deconvolver and "nterms=2". We adopted a robust value of $r = 0.5$, which is an intermediate value between the angular resolution and S/N, and also consistent with the analysis made by \citep{Ribas2023}. Additionally, we applied $r=-0.5$ to explore the disk with a better angular resolution ($\sim$5 au at 98 PC), thus observing small scale substructures. The resulting beam sizes are 0.09" $\times$ 0.07" at 0.89 mm and 0.12" $\times$ 0.10" at 1.3 mm for $r = 0.5$; and 0.07" $\times$ 0.05" at 0.89 mm and 0.11"$\times$ 0.09" at 1.3 mm for $r = -0.5$. The estimated rms for Band 7 are $4.8\times10^{-5}$ [Jy/beam] for $r=0.5$ and $5.8\times10^{-5}$ [Jy/beam] for $r=-0.5$ and for Band 6 are $4.1\times10^{-5}$ [Jy/beam] for $r=0.5$ and $3.9\times10^{-5}$ for $r=-0.5$ (see Table \ref{tab:continuum parameters}). We only used selfcalibrated extended configurations to produce the continuum images and deprojected brightness profiles. To estimate continuum fluxes, we applied aperture photometry on these images with a 5-$\sigma$ mask. The measured continuum fluxes are 388 $\pm$ 39 mJy at 0.89 mm and 168 $\pm$ 17 mJy at 1.3 mm. The maximum recoverable scales ($\theta_{\rm MRS}$), based on the most extended antenna configuration we used in this study (with a minimum baseline of 92 m in Band 6 and 41 m in Band 7, see Table \ref{tab:observation log}), are 2.7" and 1.8" at 0.89 mm and 1.3 mm, respectively\footnote{See \href{https://almascience.nrao.edu/documents-and-tools/cycle12/alma-technical-handbook}{ALMA Technical Handbook}, Eq. 3.28 and Sect. 7.2}. Both values are larger than the observed angular size of the disk. In Fig. \ref{fig:Continuum B7 ext selfcal r-05}, we present the Band 7 image with $r=-0.5$ on both linear and logarithmic scales. A ring structure at 0.45-0.50" or $\sim$45-50 au (assuming d = 98 pc) is detected.

\begin{table*}
\caption{\label{tab:continuum parameters} Continuum parameters of the extended configurations in Band 7 and Band 6 used in this study.} 
\renewcommand{\arraystretch}{1.31} 
\begin{center} 
\small 
\begin{tabular}{c|c|cccc} 
\hline 
\hline 
 ALMA                  &  ALMA   & Robust value (r) & Beam  &    RMS      &  Peak S/N  \\ 
 Project Code          & Band  &     &    &   ($\mu$Jy/beam)       &    \\ 
\hline 
2017.1.01167.S         & 6     &   0.5  &  0.12"x0.10"  &  41  &   320  \\ 
                       & 6     &  -0.5  &  0.11"x0.09"  & 39   &   320  \\ 
\hline 
2021.1.01205.S         & 7     &  0.5   &  0.09"x0.07"  &  48  &  394   \\ 
                       & 7     & -0.5   &  0.07"x0.05"   & 58  &  141  \\

\hline 
\end{tabular} 
\end{center} 
\end{table*}

\subsubsection{Continuum radial profiles}
\label{sec:3.1.2}

The radial profile of a circumstellar disk carries valuable information on the disk morphology, therefore revealing substructures such as rings and gaps. Additionally, it also provides the intensity distribution, which will further infer the spatial distribution of dust grains and gas molecules in the disk.

\begin{figure}
\includegraphics[width=\columnwidth]{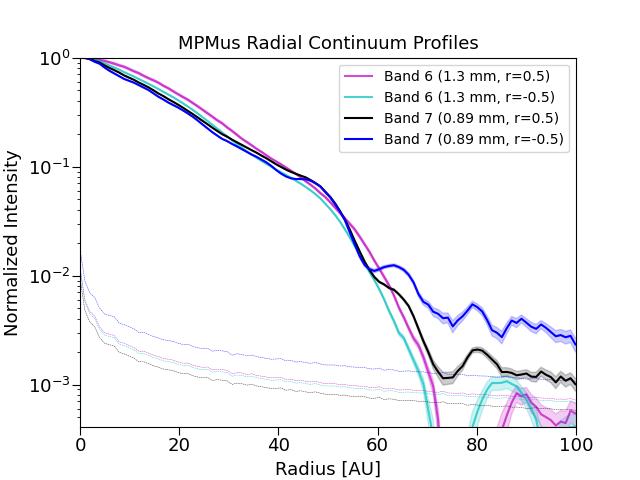}
\caption{Continuum deprojected radial profiles of MP Mus using selfcalibrated extended configurations for Band 6 (2017.1.01167.S) and Band 7, with robust values of 0.5 and -0.5. Their respective 1$\sigma$ uncertainties are presented as shades around the solid lines. The faint dotted lines in the bottom part of the figure correspond to $5\sigma$ detection limits from the rms, calculated for each dataset and each radial bin.}
\label{fig:Radial profile (cont)}
\end{figure}

The disk inclination angle (i) and position angle (PA) of MP Mus were measured as $32^\circ$ and $10^\circ$ based on a 2D Gaussian fit to the 0.89 mm dust continuum in Band 7, which is in total agreement with estimations made from Band 6 at 1.3 mm dust continuum \citep{Ribas2023} and scattered light observations \citep[e.g.,][]{Avenhaus2018}. To confirm the existence of the substructures in the disk, we deprojected the continuum image adopting the aforementioned disk inclination and PA. We then extracted the azimuthal average radial profile, as the average intensity within concentric rings centered at the peak intensity of the source. The central location is also estimated by the 2D Gaussian fit. These profiles are presented in Fig. \ref{fig:Radial profile (cont)}.

The continuum profiles are extended until $\sim$60 au in Band 7, a similar result to that in \citep{Ribas2023}. While Band 6 profiles are apparently smooth, in Band 7, some structures are seen. With a robust value of $r=-0.5$ in the Band 7 profile, we discovered what seems to be a little bump at $\sim$50 au that could correspond to the ring structure observed in Fig. \ref{fig:Continuum B7 ext selfcal r-05} since they are at the same location at $\sim 45-50$ au. Moreover, this bump is in total agreement with the feature found by \cite{Ribas2023} in Band 6. In addition, from the Band 7 profiles, it seems that a second bump emerged at $\sim 60$ au, but this is a very low S/N feature.

\subsubsection{Frank radial profile} 
\label{sec:3.1.3}

\begin{figure}
\includegraphics[width=\columnwidth]{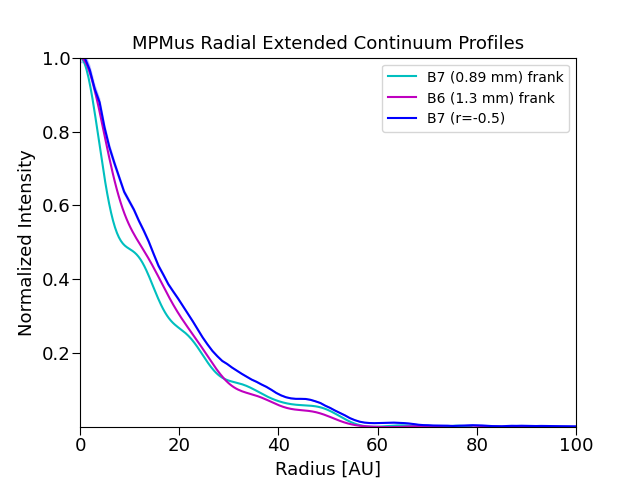}
\caption{Radial profiles of the continuum for MP Mus using selfcalibrated extended configurations in Band 7 and frank radial profile reconstruction from the visibility of Band 6 (from \citep{Ribas2023}) and Band 7. }
\label{fig:Frank profiles}
\end{figure}

Here, we used the {\sc frankenstein} software \cite[Frank;][]{Jennings2020} to construct a 1D radial profile from the visibilities of the continuum in Band 7, to perform an in depth study of the MP Mus. Frank calculates super resolution radial profiles of protoplanetary disks assuming azimuthal symmetry, an assumption that is in agreement with previous observations of MP Mus \citep{Avenhaus2018}. Using a fast ($<$1 min) Gaussian process, the Frank software fits the visibilities directly and reconstructs a 1D radial brightness profile nonparametrically. The inclination and position angle values derived from Frank are consistent with the previously adopted results. We then compare it with the one previously calculated in Band 6 by \cite{Ribas2023} and with the continuum radial profiles (see Fig. \ref{fig:Frank profiles}).

From the reconstruction of Frank profiles, we noticed that the bump at $\sim$45-50 au is present, which is in total agreement with the continuum image and the continuum radial profile in Band 7.

\subsubsection{Dust continuum versus $^{12}$CO profiles}
\label{sec:3.1.4}

The micrometer sized dust grains that reflect starlight are distributed out to greater distances than the larger particles responsible for the millimeter continuum \citep[e.g.,][]{Garufi2018, Ansdell2018}. Also, models of transitional disks show that small grains extend to larger radii than large grains (e.g., \cite{Villenave2019}). All of this is expected since small grains are predicted to be well coupled with the gas. On the other hand, large grains are much less decoupled from the gas and under the effects of radial drift, settle to the disk midplane \citep{Barriere-Fouchet2005}. 

\begin{figure}
\includegraphics[width=\columnwidth]{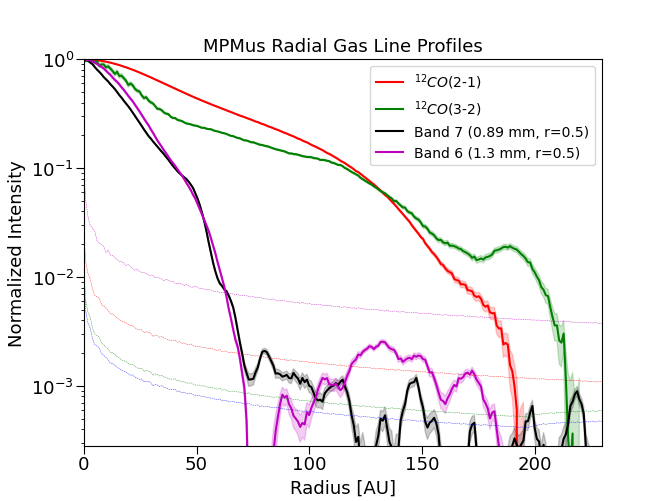}
\caption{Continuum radial profiles compared with $^{12}$CO radial profiles, with the same label definition as Fig. \ref{fig:Radial profile (cont)}.}
\label{fig:12CO_continuo}
\end{figure}

Moreover, if we look at the $^{12}$CO line profiles in Fig. \ref{fig:12CO_continuo}, where smaller dust grains sizes are coupled to the gas, the disk is much more extended than $\sim$60 au (the millimeter dust grains extension), reaching a disk extension of $\sim$110 au for $^{12}CO(2-1)$ and $\sim$120 au for $^{12}CO(3-2)$, which results in total agreement with the relation found by \citet{Andrews2020} of $R_{\rm CO} \geq 2R_{mm}$ based on 0.9 mm data from \citet{Oberg2011}, \citet{Simon2017}, \citet{Ansdell2018}, and \citet{Facchini2019}.

\subsection{Parameters calculation}
\label{sec:3.2}

\subsubsection{Disk masses}
\label{sec:3.2.1}

Submillimeter observations are ideal for estimating disk masses \citep[see examples in][]{Andrews&Williams2005}. At a scale much larger than 10 au, the continuum emission is assumed to be optically thin; therefore, the observed flux, $F_{\nu}$, is directly related to the disk mass. Here, we use the following relation to estimate the disk dust mass \citep{Hildebrand1983}:

\begin{equation}
M_{\rm dust} \simeq  \frac{F_{\nu}d^{2}}{\kappa_{\nu}B_{\nu}(T)}
\end{equation}
where $d = 98 pc $ is the distance to MP Mus from Gaia DR3. We assume $B_\nu \approx 2\nu^{2}kT/c^{2}$ since the Planck function is close to the Rayleigh-Jeans regime, a temperature of $T = 20 K$ and $\kappa_{\nu} \sim 3cm^{2}g^{-1}$ as the dust opacity at 0.89 mm \citep{Ricci2012A} and 1.3 mm \citep{Andrews2011}. This results in $(8.53\pm0.85)\times10^{-5} M_{\odot}$ or $28.4\pm2.8 M_{\oplus}$ at 0.89 mm and $(7.88\pm0.79)\times10^{-5} M_{\odot}$ or $26.3\pm2.6 M_{\oplus}$ at 1.3 mm as upper limits.

In addition, by assuming that MP Mus has properties similar to a median disk in Taurus, we use the linear relations derived by \cite{Cieza2008} to estimate the total disk mass from the observed flux:

\begin{equation}
M_{\rm disk} = 1.7 \times 10^{-1} \left [  \frac{F_{\nu}(1.3mm)}{mJy} \times \left (\frac{d}{140pc} \right )^{2} \right ] M_{J}
\end{equation}

\begin{equation}
M_{\rm disk} = 8.0 \times 10^{-2} \left [ \frac{F_{\nu}(0.85mm)}{mJy} \times \left (\frac{d}{140pc} \right )^{2}  \right ]M_{J}
\end{equation}

These relations come from modeling the IR and submillimeter SED of observations presented by \citet{Andrews&Williams2005, Andrews&Williams2007}, with a disk temperature of $T = 20 K$, which is the median disk calculated by \citet{Andrews&Williams2005}, and a dust opacity of $\kappa_{\nu} = 10[\nu/1200 GHz] \, cm^{2}g^{-1}$ that simply assumes a gas-to-dust mass ratio of 100 \citep{Williams&Cieza2011, Cieza2016}. These results on total disk masses of 15.43 $\pm$ 1.55 $M_{Jup}$ at Band 7 and of 14.2 $\pm$ 1.44 $M_{Jup}$ at Band 6. 

However, dust opacities at millimeter wavelengths have a great dependence on grain size distribution, structure, and composition of the disk, therefore, their values still remain highly uncertain \citep{Birnstiel2018, Birnstiel2024}.


\subsubsection{Spectral index}  
\label{sec:3.2.2}

In protoplanetary disks, the millimeter spectral index ($\alpha_{mm}$) has a strong dependence on the grain size distribution and, therefore, it is a widely used parameter to investigate grain growth in circumstellar disks among several star forming regions \citep[e.g.,][]{Ricci2012, Ribas2017, Ansdell2018, Tazzari2021}. 

At millimeter wavelengths, the thermal emission of a disk, integrated over its surface, is dominated by the emission of the outer, optically thin region. By assuming that the dust emission from the disk is optically thin and in the Rayleigh-Jeans limit regime (unless the disk would be abnormally cold), the wavelength dependence of the integrated flux is expressed as $F_{\nu} \propto \nu^{\beta_{mm}+2}$, where $\beta_{mm}$ is the dust opacity index. Thus, we can fit the observed submillimeter and millimeter SED between two frequencies with $F_{\nu 1}/F_{\nu 2} = (\nu_{1}/\nu_{2})^{\alpha_{mm}}$, where the sub-mm/mm spectral index is related to the dust opacity index by $\alpha_{mm} = \beta_{mm} + 2$.

\begin{figure*}
\centering
\includegraphics[width=18.5cm]{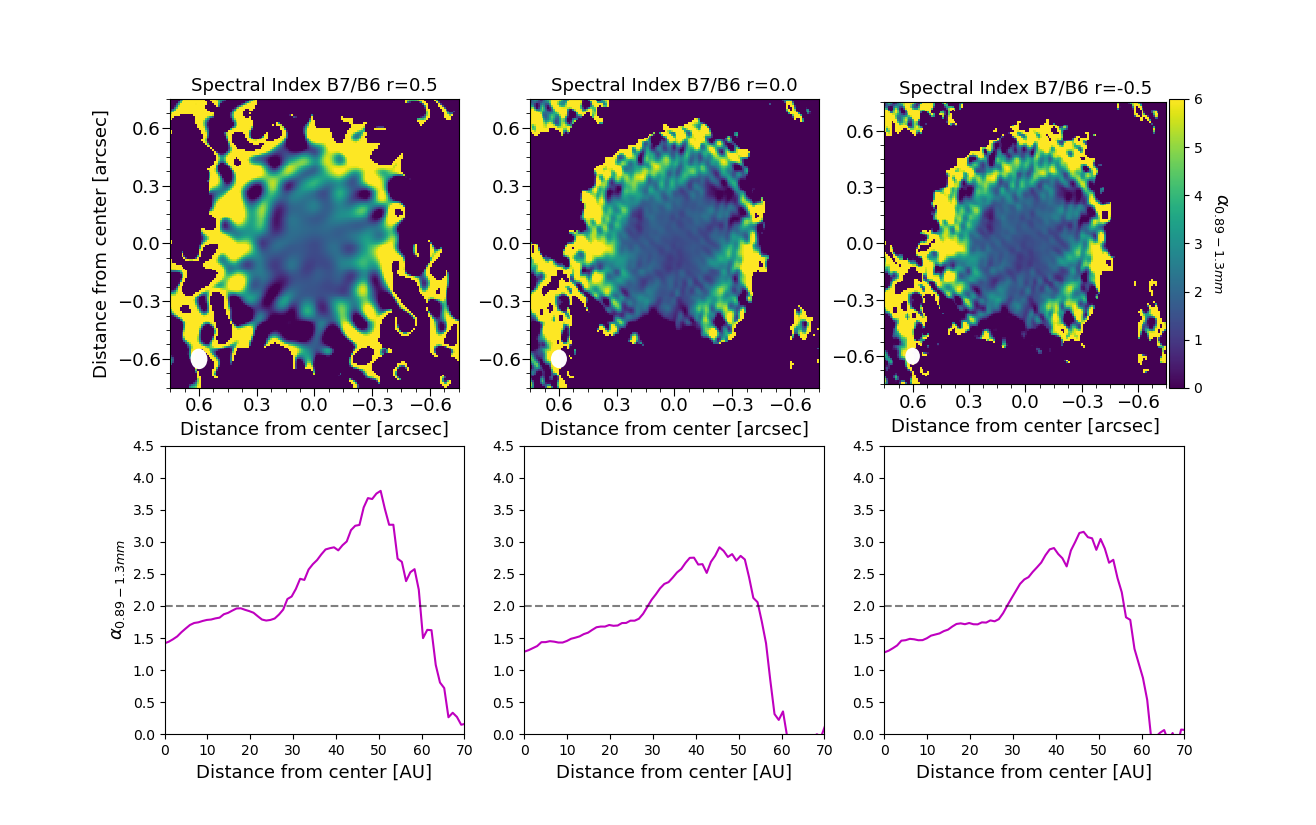}
\caption{\textit{Top:} Spectral index maps of MP Mus between 0..89 mm and 1.3 mm, with different Briggs robust values: 0.5, 0.0, and -0.5. \textit{Bottom:} Deprojected $\alpha_{0.89-1.3mm}$ radial profiles. The horizontal dashed line represents the Rayleigh-Jeans limit ($\alpha_{mm} = 2$).}
\label{fig:spectral index}
\end{figure*}

We calculated a spectral index between Band 6 and Band 7 as $\alpha_{0.89-1.3 mm} = 2.2 \pm 0.3$ by using the integrated fluxes values calculated in Sect. \ref{sec:3.1}. However, this value only represents an average spectral index in the disk, without the information of the spatial distribution. In addition, based on more recent millimeter observations at different wavelengths of several protoplanetary disks, the opacity index and, by extension, the spectral index, may change with the radius \citep{Guilloteau2011, Pinilla2014, Birnstiel2024}. In order to study the spatial variation of $\alpha$, we combined 0.89 mm and 1.3 mm observations to produce a spectral index map. To join the data, we set a common phase center estimated by a 2D Gaussian fit, and then we used the \texttt{tclean} task with \texttt{mtmfs} deconvolver, \texttt{nterms}=2, which internally sets a common beam size between the two visibility data sets. We also probe with different \texttt{robust} values (0.5, 0.0, and -0.5). The resulting \texttt{alpha} images were used as the spectral index map, and it is shown in Fig. \ref{fig:spectral index} along with its deprojected radial profiles. 
As expected, the spectral index ($\alpha_{0.89-1.3 mm}$), is variable through the disk, increasing with radius, starting with a value of $\sim$1.2 at the inner radius and finishing with a value of $\sim$4 for r = 0.5. For a further discussion, see Sect. \ref{sec:4.2}.

\section{Discussion}
\label{sec:4}

\subsection{Millimeter substructure in Band 7}
\label{sec:4.1}

 With the new dataset with high angular resolution in Band 7 ($\sim$5 au), we are able to observe a small outer ring that has been revealed at $\sim$50 au from the continuum emission at 0.89 mm. However, there are still no signs of an inner cavity. 
 
 In previous works at millimeter wavelengths, MP Mus appeared to be a smooth disk, although it has a ring structure in scattered light \citep{Avenhaus2018}. \cite{Ribas2023} reached a similar angular resolution at 1.3 mm ($\sim$4 au) as our Band 7 data, observing a smooth disk with the possibility of a barely resolved outer ring at $\sim$50 au. This is an interesting scenario, where the angular resolutions are similar in both bands and the structure is revealed at 0.89 mm observations, confirming the features found at 1.3 mm by \citep{Ribas2023}, which may be more pronounced at Band 7 since it has a larger S/N emission than that at Band 6. 
 
\begin{figure}
	\includegraphics[width=\columnwidth]{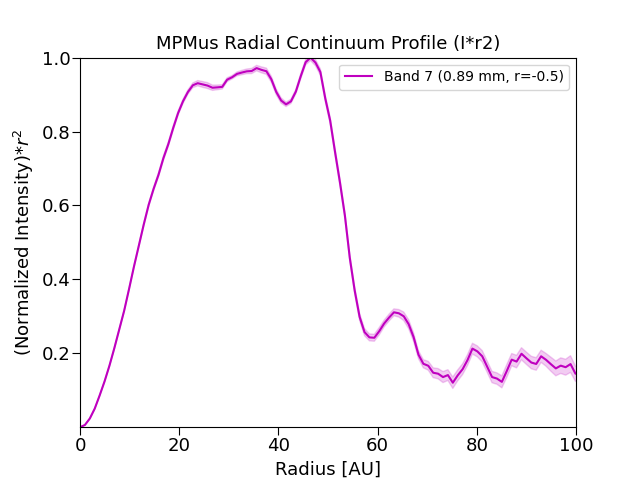}
\caption{Radial brightness profile of Band 7 continuum of MP Mus multiplying by $r^{2}$ to emphasize the feature. A robust value of $r=-0.5$ was used for this profile.}
    \label{fig:Intensity_r2}
\end{figure}

 To emphasize the structure, we made a brightness profile multiplying the brightness intensity by $r^{2}$, shown in Fig. \ref{fig:Intensity_r2}, where a peak is shown between 45-50 au, which matches the observed outer ring in the continuum image at 0.89 mm.
 
\begin{figure*}
\centering
\includegraphics[width=18.5cm]{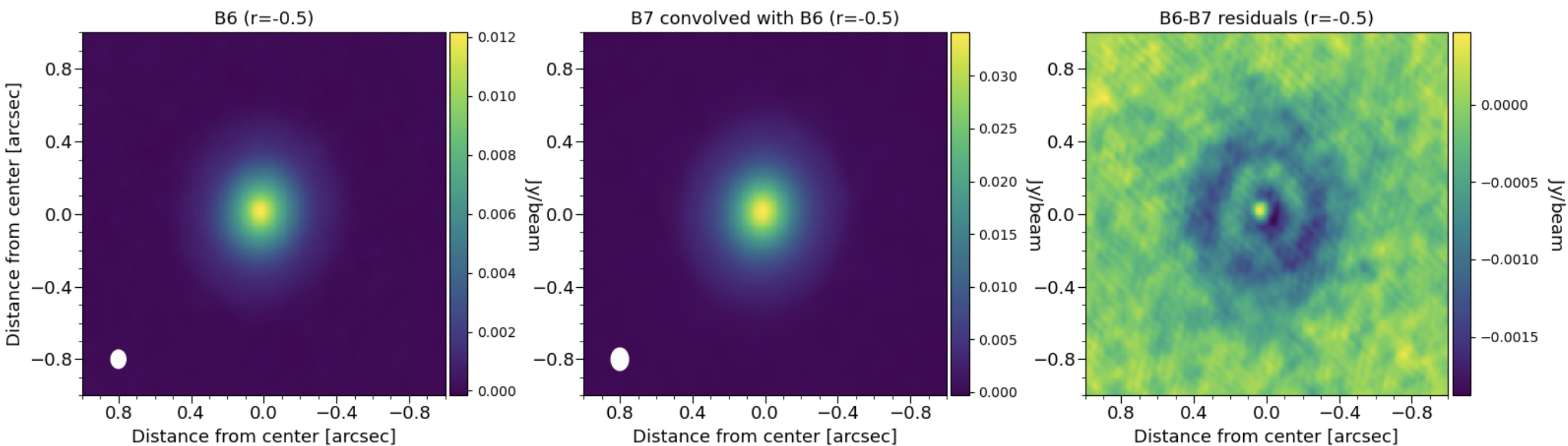}
\caption{Continuum images of MP Mus. The beam sizes are shown in the lower left corner of each image. \textit{Left:} Band 6 continuum of MP Mus with r = -0.5 and a beam size of 0.11" $\times$ 0.09". \textit{Right:} Band 7 continuum of MP Mus convolved with the beam of Band 6 and r = -0.5. The beam size is 0.13" $\times$ 0.1". }
\label{fig:B7_convolved_B6}
\end{figure*}

To understand the reasons why we are not able to observe the same structure in the 1.3 mm continuum, we convolved the 0.89 mm continuum image with the beam size of the 1.3 mm continuum image and then corrected the images to a common phase center, estimated by a 2D Gaussian fit. The resulting image, along with its residual image, are shown in Fig. \ref{fig:B7_convolved_B6}. The ring structure seems to disappear with the convolution to the larger beam size, suggesting that there is not enough angular resolution to resolve the structure in our 1.3 mm image. However, the ring structure is clearly presented in the residual image (see the right panel of Fig. \ref{fig:B7_convolved_B6}).

In addition, \cite{Pinilla2014} compared data from \cite{Ricci2012A}, who presented millimeter fluxes and integrated spectral indices of $\sim$50 classical disks in Taurus, Ophiuchus, and Orion star forming regions with data of transitional disks in Taurus, Perseus, Lupus, Ophiuchus, and, among others (references therein). They found that the mean value of the integrated spectral index for a typical protoplanetary disk corresponds to a value of $\overline{\alpha}^{PD}_{mm} = 2.20 \pm 0.07$ in comparison to a value of $\overline{\alpha}^{TD}_{mm} = 2.70 \pm 0.13$ for transitional disks. Also, the values estimated on previous works are consistent with both, classical and transitional disks: \cite{Cortes2009} found a value of ${\alpha}_{3mm-12cm} = 2.4 \pm 0.1$ and \cite{Ribas2023} found values of ${\alpha}_{0.89-1.3mm} = 2.4 \pm 0.3$, ${\alpha}_{1.3-2.2mm} = 2.12\pm 0.11$ and ${\alpha}_{0.89-2.2mm} = 2.25 \pm 0.13$. From the spectral index estimated for MP Mus (${\alpha}_{0.89-1.3mm} \sim 2.2$), its value is more concordant with classical protoplanetary disks, added to the fact that it does not present a large inner cavity, a necessary feature for MP Mus to be classified as a transitional disk.

\subsection{Indicators of grain growth and radial drift on the disk}
\label{sec:4.2}

As we mentioned before, the spectral index has been widely used to characterize the dust properties in protoplanetary disks, especially to constrain the maximum particle sizes in the submillimeter to centimeter range. The opacity spectral index, $\beta$, is sensitive to the maximum grain size of the disk, $a_{max}$ \citep[e.g.,][]{NattaTesti2004, Draine2006, Testi2014}. When dust grains reach millimeter or larger sizes ($a_\text{max}$ $>> \lambda$), the absolute value of dust opacity, $\kappa_{\lambda}$, decreases, $\beta$ becomes smaller ($\beta \sim$ 0) and the emission follows the shape of a blackbody ($\alpha_{mm} \sim 2$). Alternatively, smaller dust grains ($a_\text{max}$ $<< \lambda$, e.g., ISM like grains) have $\beta \sim$ 2. Another possible explanation for low values of $\beta$ is regions of high optical depth, where \cite{Ricci2012} has shown that this scenario would only be possible for the brightest and most massive disk. Thus, low values of $\beta$, and therefore, $\alpha_{mm}$ are indicative of significant grain growth, high optical depth, or a combination of these two processes \citep{Birnstiel2024}. From our derived spectral index $\alpha_{0.89-1.3mm}$, we found a range of $\beta$ between: 0.0-0.5.

On the other hand, compact disks at millimeter wavelengths are expected from radial drift models, where dust pebbles drift toward (local) pressure maxima present in the gas structure \citep[e.g.,][]{Weidenschilling1977}, grain growth of the millimeter grains, and also due to line optical depth. These effects predict that millimeter continuum emission should be much more compact than the emission of gas tracers, particularly of the CO tracer \citep[e.g.,][]{Facchini2017, Trapman2019}. 

\cite{BirnstielAndrews2014} demonstrate that the combined effects of radial drift and (viscous) gas drag naturally produce a sharp outer edge in the dust distribution, where this edge feature forms before grain growth has made much progress in the outer disk, and remains over longer timescales. Also, the average dust-to-gas mass ratio of the disk should be lower than the canonical 1\%, with the inner regions locally showing a higher value.

Furthermore, $\alpha_{mm}$ is variable and increases its value as a function of the radius. In Fig. \ref{fig:spectral index}, we observed a value variation of $\alpha_{0.89-1.3mm}$ between 1.2 and 3.0-4.0 depending on the robust value we used, where the blackbody limit is reached at $\sim$30 au ($\alpha = 2$). Since the spectral index is lower ($\alpha_{mm}<$2) at the inner radius, it suggests that this part of the disk is optically thick, the information about dust sizes is lost and we are observing an elevated surface of the disk with the continuum emission at 1.3 mm and 0.89 mm, which could hide some structures in the midplane of the disk. For larger radii, where $\alpha_{mm}>$2 and $\beta$ remain in lower values (between 30-40 au), it could be possible that grain growth takes part as an effective process. In addition, in regions where dust growth is effective, assuming that dust opacity is dominated by absorption is far from reality; instead, for those large particles, the dust opacity should indeed be dominated by scattering \citep[e.g.,][]{Birnstiel2018, Sierra2019}.

Since the inner part of the MP Mus disk is optically thick, we only have information on the surface, indicating a larger population of grains than the observation, which could be translated into a higher dust-to-gas ratio. Also, MP Mus shows a compact disk in millimeter continuum ($\sim$ 60~au) in contrast with gas tracers such as $^{12}$CO lines, reaching a radius of $\sim 110-120$~au (see appendix \ref{sec:C}). Moreover, we obtained an estimated ratio of $R_{gas}/R_{dust} \sim 2$, a value that is not large enough to demonstrate dust evolution, which requires an $R_{gas}/R_{dust} > 4$ \citep{Trapman2019}. Thus, it is likely that the radial drift is the most dominant process. However, identifying the dust evolution process from $R_{gas}/R_{dust}$ requires modeling the disk structure, including the total CO content.

\subsection{Comparison with disk opacity theory}
\label{sec:4.3}

\begin{figure}
	\includegraphics[width=\columnwidth]{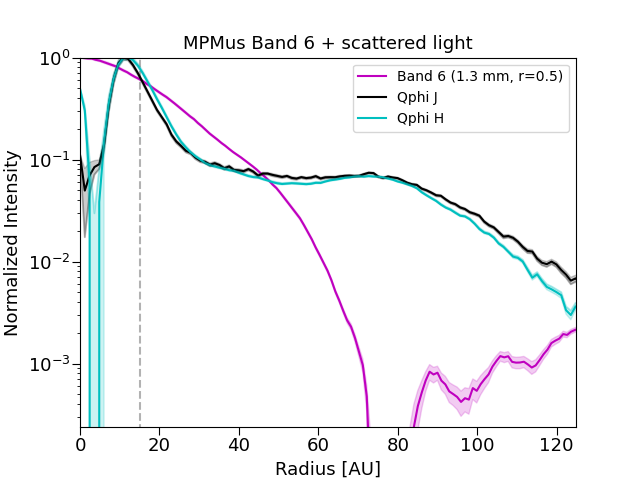}
\caption{Radial brightness profile of J and H Band from Sphere at scattered light in comparison with the Band 6 continuum with a robust value of $r=0.5$. The vertical dashed line corresponds to the coronograph limit.}
    \label{fig:sphere-mm}
\end{figure}

Substructures in the disk of MP Mus have been identified in the scattered light images from SPHERE \citep{Avenhaus2018}, with a depression of its intensity between 25-75 au. However, on the ALMA Band 6 continuum image, there were no indications of substructures within the 80 au of the disk \citep{Ribas2023}, except for a feature at $\sim$50 au. In Fig. \ref{fig:sphere-mm}, we compare the radial profiles of the near-IR scattered light (1.65 $\mu$m) that shows a peak of intensity at ~80 au with ALMA Band 6 continuum (1.3 mm), where the profile is extended until $\sim$60 au.  

\citet{Birnstiel2015} proposed that the inconsistency between the near infrared and submillimeter observed disk profiles may be an effect of dust opacity due to dust evolution. Briefly, in the distribution of particle sizes as a function of radius, there is a disk region where the population of small grains is not replenished and fragmentation is inefficient. Hence, the turbulent diffusion cannot extend there, resulting in short dust growth and drift timescales, leading to the presence of smaller dust grains \citep[see Fig. 1 of][]{Birnstiel2015}. Additionally, its predictions in the radial intensity profiles between scattered light and millimeter continuum could be in agreement with our observations at (25-50 au), where a dip in the scattered light is observed, whilst there is an absence in millimeter observations \citep[see Fig. 3 of][]{Birnstiel2015}. However, the ring is detected at $\sim$(80 au), and at this radius, we have no signal from millimeter observations. Moreover, we detected a ring at $\sim 50$ au, whose results are incompatible with the predictions for this effect, that it is based on the fact that there is no structure at 1.3 mm.

\section{Conclusions}
\label{sec:5}

Protoplanetary disks are the birthplace of planetary systems, offering a unique view of the mechanisms of planet formation and disk evolution. Transitional disks, characterized by their gaps and rings, represent an intermediate stage between gas rich primordial disks and debris dominated systems. In this work, we studied the disk structure of MP Mus, as one of the closest disks to our solar system. At a distance of 97.8 pc, MP Mus is an excellent target for investigating detailed grain growth and disk evolution procedures. Previous studies of MP Mus revealed disk structures in the near-IR (due to scattering light) but a lack of resolved detections in the submillimeter data.

Here, we presented new observations of MP Mus obtained with ALMA Band 7 at 0.89 mm. These observations provide a higher spatial resolution of $\sim$5 au that complements previous Band 6 data at 1.3 mm which reaches a resolution of $\sim$4 au, allowing us to probe the disk morphology in greater detail. The spectral index ($\alpha_{mm}$) was mapped by combining multiband ALMA dust continuum data, enabling us to investigate the radial distribution of grain sizes and their implications for disk opacity and grain growth processes. Our key findings include:

\begin{itemize}
\item{Detection of a ring structure}: A ring was confirmed at $\sim 50$~au at 0.89 mm dust continuum from the ALMA Band 7 data, a feature that was not observed in our Band 6 data due to its low spatial resolution.

\item{Extended gas disk}: A gas disk was observed to extend up to $\sim 120$~au in the $^{12}CO$ tracer, significantly beyond the dust disk ($\sim60$ au), consistent with theoretical predictions on the large gas disk radii due to radial drift. However, it can also be a contribution to dust evolution, but models for disk structure that include the total CO content are required.

\item{Spectral index variation}: The radial variation of the spectral index in Band 6 and Band 7 showed a significant increase from $\sim 1.2$ in the inner disk to $\sim 3.0-4.0$ in the outer radius ($\sim$55 au), depending on the robust value we used. This gradient reflects the changing optical depth and grain size distribution, suggesting active processes such as radial drift and dust trapping.

\item{Implications for grain growth}: The detection of substructures at shorter wavelengths and their absence in Band 6 implies that millimeter sized grains dominate the emission in the observed regions, providing evidence of ongoing grain growth and evolution.

\item{Disk classification}: Putting together the relation between spectral index and type of disks, and that there is no evidence of a large inner cavity in the disk, we cannot classify MP Mus as a transitional disk. However, the ring structure detected and the indications for the radial drift process suggest an interesting stage of evolution in the disk.
\end{itemize}
 
In summary, our results underline the complex interplay of physical processes shaping disks such as MP Mus. The presence of a ring structure at 50~au and the radial variation of the spectral index provide critical information about disk clearing mechanisms and the initial stages of planet formation. Future theoretical modeling, coupled with more observational data (e.g., JWST and ALMA), will be the key to further understanding the dynamics and evolution of such systems.

\begin{acknowledgement}

The authors thank Professor Ruobing Dong for his kind referee report on this study. We have improved the manuscript under his suggestions. The authors thank Dr. James Miley for his guidance on the {\sc frankenstein} software. A.A. acknowledges support through a Fellowship for National PhD students from ANID, grant number 21212094. A.A. acknowledges suppor from ANID, Millennium Science Initiative, via the Núcleo Milenio de Formación Planetaria (NPF). C.C. acknowledges support by ANID BASAL project FB210003. ZG is supported by the ANID FONDECYT Postdoctoral program No. 3220029. ZG is supported by the China-Chile Joint Research Fund (CCJRF No.2301) and the Chinese Academy of Sciences South America Center for Astronomy (CASSACA) Key Research Project E52H540301. CCJRF is provided by the CASSACA and established by the National Astronomical Observatories, Chinese Academy of Sciences (NAOC), and Chilean Astronomy Society (SOCHIAS) to support China-Chile collaborations in astronomy. A.R. has been supported by the UK Science and Technology Facilities Council (STFC) via the consolidated grant ST/W000997/1 and by the European Union’s Horizon 2020 research and innovation programme under the Marie Sklodowska-Curie grant agreement No. 823823 (RISE DUSTBUSTERS project). S.P. acknowledges support from FONDECYT grant 1231663 and funding from ANID – Millennium Science Initiative Program – Center Code NCN2024\_001. L.A.C. acknowledges support from the Millennium Nucleus on Young Exoplanets and their Moons (YEMS), ANID - NCN2021 080 and NCN2024 001. L.A.C. acknowledges support from ANID, FONDECYT Regular grant number 1241056, Chile. W.G. acknowledges that partial support for this work was provided by Vanderbilt University’s First Center for Autism \& Innovation. A.S.M. acknowledges support from ANID / Fondo 2022 ALMA / 31220025. This work was funded by ANID, Millennium Science Initiative, AIM23-0001. This paper makes use of the following ALMA data: ADS/JAO.ALMA 2017.1.01419.S. ALMA is a partnership of ESO (representing its member states), NSF (USA), and NINS (Japan), together with NRC (Canada), MOST and ASIAA (Taiwan), and KASI (Republic of Korea), in cooperation with the Republic of Chile. The Joint ALMA Observatory is operated by ESO, auI/NRAO and NAOJ.
\end{acknowledgement}

\bibliographystyle{aa}
\bibliography{biblio}

\begin{thebibliography}{78}
\expandafter\ifx\csname natexlab\endcsname\relax\def\natexlab#1{#1}\fi

\bibitem[{{Andrews}(2020)}]{Andrews2020}
{Andrews}, S.~M. 2020, \araa, 58, 483

\bibitem[{{Andrews} {et~al.}(2018){Andrews}, {Huang}, {P{\'e}rez}, {Isella}, {Dullemond}, {Kurtovic}, {Guzm{\'a}n}, {Carpenter}, {Wilner}, {Zhang}, {Zhu}, {Birnstiel}, {Bai}, {Benisty}, {Hughes}, {{\"O}berg}, \& {Ricci}}]{Andrews2018}
{Andrews}, S.~M., {Huang}, J., {P{\'e}rez}, L.~M., {et~al.} 2018, The Messenger, 174, 19

\bibitem[{{Andrews} \& {Williams}(2005)}]{Andrews&Williams2005}
{Andrews}, S.~M. \& {Williams}, J.~P. 2005, \apj, 631, 1134

\bibitem[{{Andrews} \& {Williams}(2007)}]{Andrews&Williams2007}
{Andrews}, S.~M. \& {Williams}, J.~P. 2007, \apj, 671, 1800

\bibitem[{{Andrews} {et~al.}(2011){Andrews}, {Wilner}, {Espaillat}, {Hughes}, {Dullemond}, {McClure}, {Qi}, \& {Brown}}]{Andrews2011}
{Andrews}, S.~M., {Wilner}, D.~J., {Espaillat}, C., {et~al.} 2011, \apj, 732, 42

\bibitem[{{Ansdell} {et~al.}(2018){Ansdell}, {Williams}, {Trapman}, {van Terwisga}, {Facchini}, {Manara}, {van der Marel}, {Miotello}, {Tazzari}, {Hogerheijde}, {Guidi}, {Testi}, \& {van Dishoeck}}]{Ansdell2018}
{Ansdell}, M., {Williams}, J.~P., {Trapman}, L., {et~al.} 2018, \apj, 859, 21

\bibitem[{{Asensio-Torres} {et~al.}(2021){Asensio-Torres}, {Henning}, {Cantalloube}, {Pinilla}, {Mesa}, {Garufi}, {Jorquera}, {Gratton}, {Chauvin}, {Szul{\'a}gyi}, {van Boekel}, {Dong}, {Marleau}, {Benisty}, {Villenave}, {Bergez-Casalou}, {Desgrange}, {Janson}, {Keppler}, {Langlois}, {M{\'e}nard}, {Rickman}, {Stolker}, {Feldt}, {Fusco}, {Gluck}, {Pavlov}, \& {Ramos}}]{Asensio-Torres2021}
{Asensio-Torres}, R., {Henning}, T., {Cantalloube}, F., {et~al.} 2021, \aap, 652, A101

\bibitem[{{Avenhaus} {et~al.}(2018){Avenhaus}, {Quanz}, {Garufi}, {Perez}, {Casassus}, {Pinte}, {Bertrang}, {Caceres}, {Benisty}, \& {Dominik}}]{Avenhaus2018}
{Avenhaus}, H., {Quanz}, S.~P., {Garufi}, A., {et~al.} 2018, \apj, 863, 44

\bibitem[{{Barri{\`e}re-Fouchet} {et~al.}(2005){Barri{\`e}re-Fouchet}, {Gonzalez}, {Murray}, {Humble}, \& {Maddison}}]{Barriere-Fouchet2005}
{Barri{\`e}re-Fouchet}, L., {Gonzalez}, J.~F., {Murray}, J.~R., {Humble}, R.~J., \& {Maddison}, S.~T. 2005, \aap, 443, 185

\bibitem[{{Birnstiel}(2024)}]{Birnstiel2024}
{Birnstiel}, T. 2024, \araa, 62, 157

\bibitem[{{Birnstiel} \& {Andrews}(2014)}]{BirnstielAndrews2014}
{Birnstiel}, T. \& {Andrews}, S.~M. 2014, \apj, 780, 153

\bibitem[{{Birnstiel} {et~al.}(2015){Birnstiel}, {Andrews}, {Pinilla}, \& {Kama}}]{Birnstiel2015}
{Birnstiel}, T., {Andrews}, S.~M., {Pinilla}, P., \& {Kama}, M. 2015, \apjl, 813, L14

\bibitem[{{Birnstiel} {et~al.}(2018){Birnstiel}, {Dullemond}, {Zhu}, {Andrews}, {Bai}, {Wilner}, {Carpenter}, {Huang}, {Isella}, {Benisty}, {P{\'e}rez}, \& {Zhang}}]{Birnstiel2018}
{Birnstiel}, T., {Dullemond}, C.~P., {Zhu}, Z., {et~al.} 2018, \apjl, 869, L45

\bibitem[{{Bouwman} {et~al.}(2008){Bouwman}, {Henning}, {Hillenbrand}, {Meyer}, {Pascucci}, {Carpenter}, {Hines}, {Kim}, {Silverstone}, {Hollenbach}, \& {Wolf}}]{Bouwman2008}
{Bouwman}, J., {Henning}, T., {Hillenbrand}, L.~A., {et~al.} 2008, \apj, 683, 479

\bibitem[{{Carpenter} {et~al.}(2005){Carpenter}, {Wolf}, {Schreyer}, {Launhardt}, \& {Henning}}]{Carpenter2005}
{Carpenter}, J.~M., {Wolf}, S., {Schreyer}, K., {Launhardt}, R., \& {Henning}, T. 2005, \aj, 129, 1049

\bibitem[{{Carrasco-Gonz{\'a}lez} {et~al.}(2016){Carrasco-Gonz{\'a}lez}, {Henning}, {Chandler}, {Linz}, {P{\'e}rez}, {Rodr{\'\i}guez}, {Galv{\'a}n-Madrid}, {Anglada}, {Birnstiel}, {van Boekel}, {Flock}, {Klahr}, {Macias}, {Menten}, {Osorio}, {Testi}, {Torrelles}, \& {Zhu}}]{Carrasco-Gonzalez2016}
{Carrasco-Gonz{\'a}lez}, C., {Henning}, T., {Chandler}, C.~J., {et~al.} 2016, \apjl, 821, L16

\bibitem[{{Cieza}(2016)}]{Cieza2016}
{Cieza}, L.~A. 2016, in IAU Symposium, Vol. 314, Young Stars \& Planets Near the Sun, ed. J.~H. {Kastner}, B.~{Stelzer}, \& S.~A. {Metchev}, 128--134

\bibitem[{{Cieza} {et~al.}(2008){Cieza}, {Swift}, {Mathews}, \& {Williams}}]{Cieza2008}
{Cieza}, L.~A., {Swift}, J.~J., {Mathews}, G.~S., \& {Williams}, J.~P. 2008, \apjl, 686, L115

\bibitem[{{Cortes} {et~al.}(2009){Cortes}, {Meyer}, {Carpenter}, {Pascucci}, {Schneider}, {Wong}, \& {Hines}}]{Cortes2009}
{Cortes}, S.~R., {Meyer}, M.~R., {Carpenter}, J.~M., {et~al.} 2009, \apj, 697, 1305

\bibitem[{{Dasgupta} {et~al.}(2025){Dasgupta}, {Cieza}, {Gonzalez Ruilova}, {Bhowmik}, {Prachi Chavan}, {Batalla-Falcon}, {Herczeg}, {Ruiz-Rodriguez}, {Williams}, {Sierra}, {Casassus}, {Guilera}, {Perez}, {Orcajo}, {Nogueira}, {Hales}, {Miley}, {Rannou}, \& {Zurlo}}]{Dasgupta2025}
{Dasgupta}, A., {Cieza}, L.~A., {Gonzalez Ruilova}, C.~I., {et~al.} 2025, arXiv e-prints, arXiv:2501.15789

\bibitem[{{Dickson-Vandervelde} {et~al.}(2021){Dickson-Vandervelde}, {Wilson}, \& {Kastner}}]{Dickson2021}
{Dickson-Vandervelde}, D.~A., {Wilson}, E.~C., \& {Kastner}, J.~H. 2021, \aj, 161, 87

\bibitem[{{Draine}(2006)}]{Draine2006}
{Draine}, B.~T. 2006, \apj, 636, 1114

\bibitem[{{Facchini} {et~al.}(2017){Facchini}, {Birnstiel}, {Bruderer}, \& {van Dishoeck}}]{Facchini2017}
{Facchini}, S., {Birnstiel}, T., {Bruderer}, S., \& {van Dishoeck}, E.~F. 2017, \aap, 605, A16

\bibitem[{{Facchini} {et~al.}(2019){Facchini}, {van Dishoeck}, {Manara}, {Tazzari}, {Maud}, {Cazzoletti}, {Rosotti}, {van der Marel}, {Pinilla}, \& {Clarke}}]{Facchini2019}
{Facchini}, S., {van Dishoeck}, E.~F., {Manara}, C.~F., {et~al.} 2019, \aap, 626, L2

\bibitem[{{Flock} {et~al.}(2015){Flock}, {Ruge}, {Dzyurkevich}, {Henning}, {Klahr}, \& {Wolf}}]{Flock2015}
{Flock}, M., {Ruge}, J.~P., {Dzyurkevich}, N., {et~al.} 2015, \aap, 574, A68

\bibitem[{{Gaia Collaboration} {et~al.}(2023){Gaia Collaboration}, {Vallenari}, {Brown}, {Prusti}, {de Bruijne}, {Arenou}, {Babusiaux}, {Biermann}, {Creevey}, {Ducourant}, {Evans}, {Eyer}, {Guerra}, {Hutton}, {Jordi}, {Klioner}, {Lammers}, {Lindegren}, {Luri}, {Mignard}, {Panem}, {Pourbaix}, {Randich}, {Sartoretti}, {Soubiran}, {Tanga}, {Walton}, {Bailer-Jones}, {Bastian}, {Drimmel}, {Jansen}, {Katz}, {Lattanzi}, {van Leeuwen}, {Bakker}, {Cacciari}, {Casta{\~n}eda}, {De Angeli}, {Fabricius}, {Fouesneau}, {Fr{\'e}mat}, {Galluccio}, {Guerrier}, {Heiter}, {Masana}, {Messineo}, {Mowlavi}, {Nicolas}, {Nienartowicz}, {Pailler}, {Panuzzo}, {Riclet}, {Roux}, {Seabroke}, {Sordo}, {Th{\'e}venin}, {Gracia-Abril}, {Portell}, {Teyssier}, {Altmann}, {Andrae}, {Audard}, {Bellas-Velidis}, {Benson}, {Berthier}, {Blomme}, {Burgess}, {Busonero}, {Busso}, {C{\'a}novas}, {Carry}, {Cellino}, {Cheek}, {Clementini}, {Damerdji}, {Davidson}, {de Teodoro}, {Nu{\~n}ez Campos}, {Delchambre}, {Dell'Oro}, {Esquej},
  {Fern{\'a}ndez-Hern{\'a}ndez}, {Fraile}, {Garabato}, {Garc{\'\i}a-Lario}, {Gosset}, {Haigron}, {Halbwachs}, {Hambly}, {Harrison}, {Hern{\'a}ndez}, {Hestroffer}, {Hodgkin}, {Holl}, {Jan{\ss}en}, {Jevardat de Fombelle}, {Jordan}, {Krone-Martins}, {Lanzafame}, {L{\"o}ffler}, {Marchal}, {Marrese}, {Moitinho}, {Muinonen}, {Osborne}, {Pancino}, {Pauwels}, {Recio-Blanco}, {Reyl{\'e}}, {Riello}, {Rimoldini}, {Roegiers}, {Rybizki}, {Sarro}, {Siopis}, {Smith}, {Sozzetti}, {Utrilla}, {van Leeuwen}, {Abbas}, {{\'A}brah{\'a}m}, {Abreu Aramburu}, {Aerts}, {Aguado}, {Ajaj}, {Aldea-Montero}, {Altavilla}, {{\'A}lvarez}, {Alves}, {Anders}, {Anderson}, {Anglada Varela}, {Antoja}, {Baines}, {Baker}, {Balaguer-N{\'u}{\~n}ez}, {Balbinot}, {Balog}, {Barache}, {Barbato}, {Barros}, {Barstow}, {Bartolom{\'e}}, {Bassilana}, {Bauchet}, {Becciani}, {Bellazzini}, {Berihuete}, {Bernet}, {Bertone}, {Bianchi}, {Binnenfeld}, {Blanco-Cuaresma}, {Blazere}, {Boch}, {Bombrun}, {Bossini}, {Bouquillon}, {Bragaglia}, {Bramante}, {Breedt},
  {Bressan}, {Brouillet}, {Brugaletta}, {Bucciarelli}, {Burlacu}, {Butkevich}, {Buzzi}, {Caffau}, {Cancelliere}, {Cantat-Gaudin}, {Carballo}, {Carlucci}, {Carnerero}, {Carrasco}, {Casamiquela}, {Castellani}, {Castro-Ginard}, {Chaoul}, {Charlot}, {Chemin}, {Chiaramida}, {Chiavassa}, {Chornay}, {Comoretto}, {Contursi}, {Cooper}, {Cornez}, {Cowell}, {Crifo}, {Cropper}, {Crosta}, {Crowley}, {Dafonte}, {Dapergolas}, {David}, {David}, {de Laverny}, {De Luise}, {De March}, {De Ridder}, {de Souza}, {de Torres}, {del Peloso}, {del Pozo}, {Delbo}, {Delgado}, {Delisle}, {Demouchy}, {Dharmawardena}, {Di Matteo}, {Diakite}, {Diener}, {Distefano}, {Dolding}, {Edvardsson}, {Enke}, {Fabre}, {Fabrizio}, {Faigler}, {Fedorets}, {Fernique}, {Fienga}, {Figueras}, {Fournier}, {Fouron}, {Fragkoudi}, {Gai}, {Garcia-Gutierrez}, {Garcia-Reinaldos}, {Garc{\'\i}a-Torres}, {Garofalo}, {Gavel}, {Gavras}, {Gerlach}, {Geyer}, {Giacobbe}, {Gilmore}, {Girona}, {Giuffrida}, {Gomel}, {Gomez}, {Gonz{\'a}lez-N{\'u}{\~n}ez},
  {Gonz{\'a}lez-Santamar{\'\i}a}, {Gonz{\'a}lez-Vidal}, {Granvik}, {Guillout}, {Guiraud}, {Guti{\'e}rrez-S{\'a}nchez}, {Guy}, {Hatzidimitriou}, {Hauser}, {Haywood}, {Helmer}, {Helmi}, {Sarmiento}, {Hidalgo}, {Hilger}, {H{\l}adczuk}, {Hobbs}, {Holland}, {Huckle}, {Jardine}, {Jasniewicz}, {Jean-Antoine Piccolo}, {Jim{\'e}nez-Arranz}, {Jorissen}, {Juaristi Campillo}, {Julbe}, {Karbevska}, {Kervella}, {Khanna}, {Kontizas}, {Kordopatis}, {Korn}, {K{\'o}sp{\'a}l}, {Kostrzewa-Rutkowska}, {Kruszy{\'n}ska}, {Kun}, {Laizeau}, {Lambert}, {Lanza}, {Lasne}, {Le Campion}, {Lebreton}, {Lebzelter}, {Leccia}, {Leclerc}, {Lecoeur-Taibi}, {Liao}, {Licata}, {Lindstr{\o}m}, {Lister}, {Livanou}, {Lobel}, {Lorca}, {Loup}, {Madrero Pardo}, {Magdaleno Romeo}, {Managau}, {Mann}, {Manteiga}, {Marchant}, {Marconi}, {Marcos}, {Marcos Santos}, {Mar{\'\i}n Pina}, {Marinoni}, {Marocco}, {Marshall}, {Martin Polo}, {Mart{\'\i}n-Fleitas}, {Marton}, {Mary}, {Masip}, {Massari}, {Mastrobuono-Battisti}, {Mazeh}, {McMillan}, {Messina}, {Michalik},
  {Millar}, {Mints}, {Molina}, {Molinaro}, {Moln{\'a}r}, {Monari}, {Mongui{\'o}}, {Montegriffo}, {Montero}, {Mor}, {Mora}, {Morbidelli}, {Morel}, {Morris}, {Muraveva}, {Murphy}, {Musella}, {Nagy}, {Noval}, {Oca{\~n}a}, {Ogden}, {Ordenovic}, {Osinde}, {Pagani}, {Pagano}, {Palaversa}, {Palicio}, {Pallas-Quintela}, {Panahi}, {Payne-Wardenaar}, {Pe{\~n}alosa Esteller}, {Penttil{\"a}}, {Pichon}, {Piersimoni}, {Pineau}, {Plachy}, {Plum}, {Poggio}, {Pr{\v{s}}a}, {Pulone}, {Racero}, {Ragaini}, {Rainer}, {Raiteri}, {Rambaux}, {Ramos}, {Ramos-Lerate}, {Re Fiorentin}, {Regibo}, {Richards}, {Rios Diaz}, {Ripepi}, {Riva}, {Rix}, {Rixon}, {Robichon}, {Robin}, {Robin}, {Roelens}, {Rogues}, {Rohrbasser}, {Romero-G{\'o}mez}, {Rowell}, {Royer}, {Ruz Mieres}, {Rybicki}, {Sadowski}, {S{\'a}ez N{\'u}{\~n}ez}, {Sagrist{\`a} Sell{\'e}s}, {Sahlmann}, {Salguero}, {Samaras}, {Sanchez Gimenez}, {Sanna}, {Santove{\~n}a}, {Sarasso}, {Schultheis}, {Sciacca}, {Segol}, {Segovia}, {S{\'e}gransan}, {Semeux}, {Shahaf}, {Siddiqui}, {Siebert},
  {Siltala}, {Silvelo}, {Slezak}, {Slezak}, {Smart}, {Snaith}, {Solano}, {Solitro}, {Souami}, {Souchay}, {Spagna}, {Spina}, {Spoto}, {Steele}, {Steidelm{\"u}ller}, {Stephenson}, {S{\"u}veges}, {Surdej}, {Szabados}, {Szegedi-Elek}, {Taris}, {Taylor}, {Teixeira}, {Tolomei}, {Tonello}, {Torra}, {Torra}, {Torralba Elipe}, {Trabucchi}, {Tsounis}, {Turon}, {Ulla}, {Unger}, {Vaillant}, {van Dillen}, {van Reeven}, {Vanel}, {Vecchiato}, {Viala}, {Vicente}, {Voutsinas}, {Weiler}, {Wevers}, {Wyrzykowski}, {Yoldas}, {Yvard}, {Zhao}, {Zorec}, {Zucker}, \& {Zwitter}}]{Gaia2024}
{Gaia Collaboration}, {Vallenari}, A., {Brown}, A.~G.~A., {et~al.} 2023, \aap, 674, A1

\bibitem[{{Garufi} {et~al.}(2018){Garufi}, {Benisty}, {Pinilla}, {Tazzari}, {Dominik}, {Ginski}, {Henning}, {Kral}, {Langlois}, {M{\'e}nard}, {Stolker}, {Szulagyi}, {Villenave}, \& {van der Plas}}]{Garufi2018}
{Garufi}, A., {Benisty}, M., {Pinilla}, P., {et~al.} 2018, \aap, 620, A94

\bibitem[{{Gregorio-Hetem} {et~al.}(1992){Gregorio-Hetem}, {Lepine}, {Quast}, {Torres}, \& {de La Reza}}]{Gregorio1992}
{Gregorio-Hetem}, J., {Lepine}, J.~R.~D., {Quast}, G.~R., {Torres}, C.~A.~O., \& {de La Reza}, R. 1992, \aj, 103, 549

\bibitem[{{Gressel} {et~al.}(2015){Gressel}, {Turner}, {Nelson}, \& {McNally}}]{Gressel2015}
{Gressel}, O., {Turner}, N.~J., {Nelson}, R.~P., \& {McNally}, C.~P. 2015, \apj, 801, 84

\bibitem[{{Grimble} {et~al.}(2024){Grimble}, {Kastner}, {Pinte}, {Sargent}, {Principe}, {Dickson-Vandervelde}, {Bel{\'e}n Aguayo}, {Caceres}, {Schreiber}, \& {Stassun}}]{Grimble2024}
{Grimble}, W., {Kastner}, J., {Pinte}, C., {et~al.} 2024, \apj, 970, 137

\bibitem[{{Guilloteau} {et~al.}(2011){Guilloteau}, {Dutrey}, {Pi{\'e}tu}, \& {Boehler}}]{Guilloteau2011}
{Guilloteau}, S., {Dutrey}, A., {Pi{\'e}tu}, V., \& {Boehler}, Y. 2011, \aap, 529, A105

\bibitem[{{Hildebrand}(1983)}]{Hildebrand1983}
{Hildebrand}, R.~H. 1983, \qjras, 24, 267

\bibitem[{{Hu} {et~al.}(2019){Hu}, {Zhu}, {Okuzumi}, {Bai}, {Wang}, {Tomida}, \& {Stone}}]{Hu2019}
{Hu}, X., {Zhu}, Z., {Okuzumi}, S., {et~al.} 2019, \apj, 885, 36

\bibitem[{{Huang} {et~al.}(2018){Huang}, {Andrews}, {Dullemond}, {Isella}, {P{\'e}rez}, {Guzm{\'a}n}, {{\"O}berg}, {Zhu}, {Zhang}, {Bai}, {Benisty}, {Birnstiel}, {Carpenter}, {Hughes}, {Ricci}, {Weaver}, \& {Wilner}}]{Huang2018}
{Huang}, J., {Andrews}, S.~M., {Dullemond}, C.~P., {et~al.} 2018, \apjl, 869, L42

\bibitem[{{Ingleby} {et~al.}(2013){Ingleby}, {Calvet}, {Herczeg}, {Blaty}, {Walter}, {Ardila}, {Alexander}, {Edwards}, {Espaillat}, {Gregory}, {Hillenbrand}, \& {Brown}}]{Ingleby2013}
{Ingleby}, L., {Calvet}, N., {Herczeg}, G., {et~al.} 2013, \apj, 767, 112

\bibitem[{{Jennings} {et~al.}(2020){Jennings}, {Booth}, {Tazzari}, {Rosotti}, \& {Clarke}}]{Jennings2020}
{Jennings}, J., {Booth}, R.~A., {Tazzari}, M., {Rosotti}, G.~P., \& {Clarke}, C.~J. 2020, \mnras, 495, 3209

\bibitem[{{Kastner} {et~al.}(2010){Kastner}, {Hily-Blant}, {Sacco}, {Forveille}, \& {Zuckerman}}]{Kastner2010}
{Kastner}, J.~H., {Hily-Blant}, P., {Sacco}, G.~G., {Forveille}, T., \& {Zuckerman}, B. 2010, \apjl, 723, L248

\bibitem[{{Kastner} \& {Principe}(2022)}]{KastnerPrincipe2022}
{Kastner}, J.~H. \& {Principe}, D.~A. 2022, in Handbook of X-ray and Gamma-ray Astrophysics, ed. C.~{Bambi} \& A.~{Sangangelo}, 49

\bibitem[{{Kuwahara} {et~al.}(2024){Kuwahara}, {Lambrechts}, {Kurokawa}, {Okuzumi}, \& {Tanigawa}}]{Kuwahara2024}
{Kuwahara}, A., {Lambrechts}, M., {Kurokawa}, H., {Okuzumi}, S., \& {Tanigawa}, T. 2024, \aap, 692, A45

\bibitem[{{Long} {et~al.}(2018){Long}, {Pinilla}, {Herczeg}, {Harsono}, {Dipierro}, {Pascucci}, {Hendler}, {Tazzari}, {Ragusa}, {Salyk}, {Edwards}, {Lodato}, {van de Plas}, {Johnstone}, {Liu}, {Boehler}, {Cabrit}, {Manara}, {Menard}, {Mulders}, {Nisini}, {Fischer}, {Rigliaco}, {Banzatti}, {Avenhaus}, \& {Gully-Santiago}}]{Long2018}
{Long}, F., {Pinilla}, P., {Herczeg}, G.~J., {et~al.} 2018, \apj, 869, 17

\bibitem[{{Mamajek} {et~al.}(2002){Mamajek}, {Meyer}, \& {Liebert}}]{Mamajek2002}
{Mamajek}, E.~E., {Meyer}, M.~R., \& {Liebert}, J. 2002, \aj, 124, 1670

\bibitem[{{McMullin} {et~al.}(2007){McMullin}, {Waters}, {Schiebel}, {Young}, \& {Golap}}]{McMullin2007}
{McMullin}, J.~P., {Waters}, B., {Schiebel}, D., {Young}, W., \& {Golap}, K. 2007, in Astronomical Society of the Pacific Conference Series, Vol. 376, Astronomical Data Analysis Software and Systems XVI, ed. R.~A. {Shaw}, F.~{Hill}, \& D.~J. {Bell}, 127

\bibitem[{{Murphy} {et~al.}(2013){Murphy}, {Lawson}, \& {Bessell}}]{Murphy2013}
{Murphy}, S.~J., {Lawson}, W.~A., \& {Bessell}, M.~S. 2013, \mnras, 435, 1325

\bibitem[{{Natta} \& {Testi}(2004)}]{NattaTesti2004}
{Natta}, A. \& {Testi}, L. 2004, in Astronomical Society of the Pacific Conference Series, Vol. 323, Star Formation in the Interstellar Medium: In Honor of David Hollenbach, ed. D.~{Johnstone}, F.~C. {Adams}, D.~N.~C. {Lin}, D.~A. {Neufeeld}, \& E.~C. {Ostriker}, 279

\bibitem[{{{\"O}berg} {et~al.}(2021){{\"O}berg}, {Guzm{\'a}n}, {Walsh}, {Aikawa}, {Bergin}, {Law}, {Loomis}, {Alarc{\'o}n}, {Andrews}, {Bae}, {Bergner}, {Boehler}, {Booth}, {Bosman}, {Calahan}, {Cataldi}, {Cleeves}, {Czekala}, {Furuya}, {Huang}, {Ilee}, {Kurtovic}, {Le Gal}, {Liu}, {Long}, {M{\'e}nard}, {Nomura}, {P{\'e}rez}, {Qi}, {Schwarz}, {Sierra}, {Teague}, {Tsukagoshi}, {Yamato}, {van't Hoff}, {Waggoner}, {Wilner}, \& {Zhang}}]{Oberg2021}
{{\"O}berg}, K.~I., {Guzm{\'a}n}, V.~V., {Walsh}, C., {et~al.} 2021, \apjs, 257, 1

\bibitem[{{{\"O}berg} {et~al.}(2011){{\"O}berg}, {Qi}, {Fogel}, {Bergin}, {Andrews}, {Espaillat}, {Wilner}, {Pascucci}, \& {Kastner}}]{Oberg2011}
{{\"O}berg}, K.~I., {Qi}, C., {Fogel}, J. K.~J., {et~al.} 2011, \apj, 734, 98

\bibitem[{{Okuzumi} {et~al.}(2016){Okuzumi}, {Momose}, {Sirono}, {Kobayashi}, \& {Tanaka}}]{Okuzumi2016}
{Okuzumi}, S., {Momose}, M., {Sirono}, S.-i., {Kobayashi}, H., \& {Tanaka}, H. 2016, \apj, 821, 82

\bibitem[{{Okuzumi} \& {Tazaki}(2019)}]{OkuzumiTazaki2019}
{Okuzumi}, S. \& {Tazaki}, R. 2019, \apj, 878, 132

\bibitem[{{P{\'e}rez} {et~al.}(2015){P{\'e}rez}, {Chandler}, {Isella}, {Carpenter}, {Andrews}, {Calvet}, {Corder}, {Deller}, {Dullemond}, {Greaves}, {Harris}, {Henning}, {Kwon}, {Lazio}, {Linz}, {Mundy}, {Ricci}, {Sargent}, {Storm}, {Tazzari}, {Testi}, \& {Wilner}}]{Perez2015}
{P{\'e}rez}, L.~M., {Chandler}, C.~J., {Isella}, A., {et~al.} 2015, \apj, 813, 41

\bibitem[{{Pinilla} {et~al.}(2014){Pinilla}, {Benisty}, {Birnstiel}, {Ricci}, {Isella}, {Natta}, {Dullemond}, {Quiroga-Nu{\~n}ez}, {Henning}, \& {Testi}}]{Pinilla2014}
{Pinilla}, P., {Benisty}, M., {Birnstiel}, T., {et~al.} 2014, \aap, 564, A51

\bibitem[{{Pinte} {et~al.}(2023){Pinte}, {Teague}, {Flaherty}, {Hall}, {Facchini}, \& {Casassus}}]{Pinte2023}
{Pinte}, C., {Teague}, R., {Flaherty}, K., {et~al.} 2023, in Astronomical Society of the Pacific Conference Series, Vol. 534, Protostars and Planets VII, ed. S.~{Inutsuka}, Y.~{Aikawa}, T.~{Muto}, K.~{Tomida}, \& M.~{Tamura}, 645

\bibitem[{{Ribas} {et~al.}(2017){Ribas}, {Espaillat}, {Mac{\'\i}as}, {Bouy}, {Andrews}, {Calvet}, {Naylor}, {Riviere-Marichalar}, {van der Wiel}, \& {Wilner}}]{Ribas2017}
{Ribas}, {\'A}., {Espaillat}, C.~C., {Mac{\'\i}as}, E., {et~al.} 2017, \apj, 849, 63

\bibitem[{{Ribas} {et~al.}(2023){Ribas}, {Mac{\'\i}as}, {Weber}, {P{\'e}rez}, {Cuello}, {Dong}, {Aguayo}, {C{\'a}ceres}, {Carpenter}, {Dent}, {de Gregorio-Monsalvo}, {Duch{\^e}ne}, {Espaillat}, {Riviere-Marichalar}, \& {Villenave}}]{Ribas2023}
{Ribas}, {\'A}., {Mac{\'\i}as}, E., {Weber}, P., {et~al.} 2023, \aap, 673, A77

\bibitem[{{Ricci} {et~al.}(2012{\natexlab{a}}){Ricci}, {Testi}, {Natta}, {Scholz}, \& {de Gregorio-Monsalvo}}]{Ricci2012}
{Ricci}, L., {Testi}, L., {Natta}, A., {Scholz}, A., \& {de Gregorio-Monsalvo}, I. 2012{\natexlab{a}}, \apjl, 761, L20

\bibitem[{{Ricci} {et~al.}(2012{\natexlab{b}}){Ricci}, {Trotta}, {Testi}, {Natta}, {Isella}, \& {Wilner}}]{Ricci2012A}
{Ricci}, L., {Trotta}, F., {Testi}, L., {et~al.} 2012{\natexlab{b}}, \aap, 540, A6

\bibitem[{{Ruge} {et~al.}(2016){Ruge}, {Flock}, {Wolf}, {Dzyurkevich}, {Fromang}, {Henning}, {Klahr}, \& {Meheut}}]{Ruge2016}
{Ruge}, J.~P., {Flock}, M., {Wolf}, S., {et~al.} 2016, \aap, 590, A17

\bibitem[{{Sch{\"u}tz} {et~al.}(2005){Sch{\"u}tz}, {Meeus}, \& {Sterzik}}]{Schutz2005a}
{Sch{\"u}tz}, O., {Meeus}, G., \& {Sterzik}, M.~F. 2005, \aap, 431, 165

\bibitem[{{Segura-Cox} {et~al.}(2020){Segura-Cox}, {Schmiedeke}, {Pineda}, {Stephens}, {Fern{\'a}ndez-L{\'o}pez}, {Looney}, {Caselli}, {Li}, {Mundy}, {Kwon}, \& {Harris}}]{SeguraCox2020}
{Segura-Cox}, D.~M., {Schmiedeke}, A., {Pineda}, J.~E., {et~al.} 2020, \nat, 586, 228

\bibitem[{{Shi} {et~al.}(2024){Shi}, {Long}, {Herczeg}, {Harsono}, {Liu}, {Pinilla}, {Ragusa}, {Johnstone}, {Bai}, {Pascucci}, {Manara}, {Mulders}, \& {Cieza}}]{Shi2024}
{Shi}, Y., {Long}, F., {Herczeg}, G.~J., {et~al.} 2024, \apj, 966, 59

\bibitem[{{Sierra} {et~al.}(2017){Sierra}, {Lizano}, \& {Barge}}]{Sierra2017}
{Sierra}, A., {Lizano}, S., \& {Barge}, P. 2017, \apj, 850, 115

\bibitem[{{Sierra} {et~al.}(2019){Sierra}, {Lizano}, {Mac{\'\i}as}, {Carrasco-Gonz{\'a}lez}, {Osorio}, \& {Flock}}]{Sierra2019}
{Sierra}, A., {Lizano}, S., {Mac{\'\i}as}, E., {et~al.} 2019, \apj, 876, 7

\bibitem[{{Simon} {et~al.}(2017){Simon}, {Guilloteau}, {Di Folco}, {Dutrey}, {Grosso}, {Pi{\'e}tu}, {Chapillon}, {Prato}, {Schaefer}, {Rice}, \& {Boehler}}]{Simon2017}
{Simon}, M., {Guilloteau}, S., {Di Folco}, E., {et~al.} 2017, \apj, 844, 158

\bibitem[{{Takahashi} \& {Inutsuka}(2014)}]{TakahashiInutsuka2014}
{Takahashi}, S.~Z. \& {Inutsuka}, S.-i. 2014, \apj, 794, 55

\bibitem[{{Tazzari} {et~al.}(2021){Tazzari}, {Testi}, {Natta}, {Williams}, {Ansdell}, {Carpenter}, {Facchini}, {Guidi}, {Hogherheijde}, {Manara}, {Miotello}, \& {van der Marel}}]{Tazzari2021}
{Tazzari}, M., {Testi}, L., {Natta}, A., {et~al.} 2021, \mnras, 506, 5117

\bibitem[{{Testi} {et~al.}(2014){Testi}, {Birnstiel}, {Ricci}, {Andrews}, {Blum}, {Carpenter}, {Dominik}, {Isella}, {Natta}, {Williams}, \& {Wilner}}]{Testi2014}
{Testi}, L., {Birnstiel}, T., {Ricci}, L., {et~al.} 2014, in Protostars and Planets VI, ed. H.~{Beuther}, R.~S. {Klessen}, C.~P. {Dullemond}, \& T.~{Henning}, 339--361

\bibitem[{{Torres} {et~al.}(2008){Torres}, {Quast}, {Melo}, \& {Sterzik}}]{Torres2008}
{Torres}, C.~A.~O., {Quast}, G.~R., {Melo}, C.~H.~F., \& {Sterzik}, M.~F. 2008, in Handbook of Star Forming Regions, Volume II, ed. B.~{Reipurth}, Vol.~5, 757

\bibitem[{{Trapman} {et~al.}(2019){Trapman}, {Facchini}, {Hogerheijde}, {van Dishoeck}, \& {Bruderer}}]{Trapman2019}
{Trapman}, L., {Facchini}, S., {Hogerheijde}, M.~R., {van Dishoeck}, E.~F., \& {Bruderer}, S. 2019, \aap, 629, A79

\bibitem[{{van der Marel} {et~al.}(2016){van der Marel}, {Cazzoletti}, {Pinilla}, \& {Garufi}}]{VanderMarel2016}
{van der Marel}, N., {Cazzoletti}, P., {Pinilla}, P., \& {Garufi}, A. 2016, \apj, 832, 178

\bibitem[{{Varga} {et~al.}(2024){Varga}, {Kastner}, {Dickson-Vandervelde}, \& {Binks}}]{Varga2024}
{Varga}, A., {Kastner}, J.~H., {Dickson-Vandervelde}, D.~A., \& {Binks}, A. 2024, \aj, 168, 251

\bibitem[{{Villenave} {et~al.}(2019){Villenave}, {Benisty}, {Dent}, {M{\'e}nard}, {Garufi}, {Ginski}, {Pinilla}, {Pinte}, {Williams}, {de Boer}, {Morino}, {Fukagawa}, {Dominik}, {Flock}, {Henning}, {Juh{\'a}sz}, {Keppler}, {Muro-Arena}, {Olofsson}, {P{\'e}rez}, {van der Plas}, {Zurlo}, {Carle}, {Feautrier}, {Pavlov}, {Pragt}, {Ramos}, {Sauvage}, {Stadler}, \& {Weber}}]{Villenave2019}
{Villenave}, M., {Benisty}, M., {Dent}, W.~R.~F., {et~al.} 2019, \aap, 624, A7

\bibitem[{{Ward} \& {Hahn}(2000)}]{Ward2000}
{Ward}, W.~R. \& {Hahn}, J.~M. 2000, in Protostars and Planets IV, ed. V.~{Mannings}, A.~P. {Boss}, \& S.~S. {Russell}, 1135

\bibitem[{{Weidenschilling}(1977)}]{Weidenschilling1977}
{Weidenschilling}, S.~J. 1977, \mnras, 180, 57

\bibitem[{{Weise} {et~al.}(2010){Weise}, {Launhardt}, {Setiawan}, \& {Henning}}]{Weise2010}
{Weise}, P., {Launhardt}, R., {Setiawan}, J., \& {Henning}, T. 2010, \aap, 517, A88

\bibitem[{{Williams} \& {Cieza}(2011)}]{Williams&Cieza2011}
{Williams}, J.~P. \& {Cieza}, L.~A. 2011, \araa, 49, 67

\bibitem[{{Wolff} {et~al.}(2016){Wolff}, {Perrin}, {Millar-Blanchaer}, {Nielsen}, {Wang}, {Cardwell}, {Chilcote}, {Dong}, {Draper}, {Duch{\^e}ne}, {Fitzgerald}, {Goodsell}, {Grady}, {Graham}, {Greenbaum}, {Hartung}, {Hibon}, {Hines}, {Hung}, {Kalas}, {Macintosh}, {Marchis}, {Marois}, {Pueyo}, {Rantakyr{\"o}}, {Schneider}, {Sivaramakrishnan}, \& {Wiktorowicz}}]{Wolff2016}
{Wolff}, S.~G., {Perrin}, M., {Millar-Blanchaer}, M.~A., {et~al.} 2016, \apjl, 818, L15

\bibitem[{{Zhang} {et~al.}(2015){Zhang}, {Blake}, \& {Bergin}}]{Zhang2015}
{Zhang}, K., {Blake}, G., \& {Bergin}, E. 2015, in IAU General Assembly, Vol.~29, 2256118

\bibitem[{{Zhu} {et~al.}(2014){Zhu}, {Stone}, {Rafikov}, \& {Bai}}]{Zhu2014}
{Zhu}, Z., {Stone}, J.~M., {Rafikov}, R.~R., \& {Bai}, X.-n. 2014, \apj, 785, 122

\bibitem[{{Zurlo} {et~al.}(2020){Zurlo}, {Cugno}, {Montesinos}, {Perez}, {Canovas}, {Casassus}, {Christiaens}, {Cieza}, \& {Huelamo}}]{Zurlo2020}
{Zurlo}, A., {Cugno}, G., {Montesinos}, M., {et~al.} 2020, \aap, 633, A119

\end{thebibliography}

\appendix
\label{sec:A}

\onecolumn
\section{Tables}

\begin{table*}[h!]
\caption{\label{tab:observation log}ALMA observation log.} 
\renewcommand{\arraystretch}{1.31} 
\begin{center} 
\small 
\begin{tabular}{c|c|cccccc} 
\hline 
\hline 
 ALMA                  & ALMA  & Date       & Time on source    & $N_{\rm ant}$   & Baselines & PWV   & Flux Calibrator\\ 
 Project Code          & Band  &            & (min)             &                 & (m)       & (mm)  &  \\ 
\hline 
2017.1.01167.S         & 6     & 2017-11-16 &  11.4   & 44              & 92 - 8283 & 1.1   & J1427-4206 \\ 
  & 6     & 2018-01-15 &    5.6  & 46              & 15 - 2387 & 1.6   & J1427-4206 \\ 
\hline 
2017.1.01419.S         & 6     & 2017-12-26 &  16.7  & 43              & 15 - 2516 & 0.34  & J1427-4206 \\ 
  & 6     & 2018-06-06 &    8.3   & 44              & 15 - 313  & 0.72   & J1427-4206 \\ 
\hline 
2021.1.01205.S         & 7     & 2021-11-20 &  24.2   & 44              & 41 - 3638 & 0.36  & J1427-4207 \\ 
  & 7     & 2022-08-19 &    6.0   & 44              & 15 - 1301 & 0.32  & J1427-4206\\
\hline 
\end{tabular} 
\end{center} 
\end{table*}

\begin{table*}[h!]
\caption{\label{tab:correlator config log} ALMA correlator configuration log.} 
\renewcommand{\arraystretch}{1.31} 
\begin{center} 
\small 
\begin{tabular}{c|c|ccccc} 
\hline 
\hline 
 ALMA & ALMA    & Central Freq.         & Bandwidth    & Channels       & Spectral lines    \\
Project Code    &    Band      & (GHz)    & (Mhz)     &     &      \\    
\hline 
2017.1.01167.S & 6 &  232.485  &  2000  &   128   &  ...  \\ 
  &   &  244.984 &  2000 &  128 & ... \\ 
  &   &  246.984 &  20000 &  128 &  ...  \\ 
  &   &  230.531 &  1875 &  960 &  $^{12}CO(2-1)$  \\ 

\hline 
2017.1.01419.S & 6 &  217.550  &  2000   &   128   &  ...  \\ 
  &   &  232.800 &  2000 &  128 & ... \\ 
  &   &  219.494 &  1875 &  3840 &  $^{13}CO(2-1), C^{18}O(2-1)$  \\ 
  &   &  230.620 &  234 &  1920 &  $^{12}CO(2-1)$  \\ 
  &   &  231.205 &  234 & 1920  &  ... \\ 

\hline 
2021.1.01205.S  & 7 & 343.609 & 2000  & 128  & ...  \\ 
  &   &  345.188 &  117 &  960 &  ... \\ 
  &   &  345.788 &  117 &  960 &  $^{12}CO(3-2)$ \\ 
  &   &  355.067 &  1875 &  3840 &  $HCN(4-3)$ \\ 
  &   &  356.722 &  938 &  3840 &  $HCO^{+}(4-3)$ \\ 
\hline 
\end{tabular} 
\end{center} 
\end{table*}

\FloatBarrier
\twocolumn

\onecolumn

\section{Line cube images}
\label{sec:B}

\begin{figure*}[h!]
\centering
\includegraphics[width=18cm]
{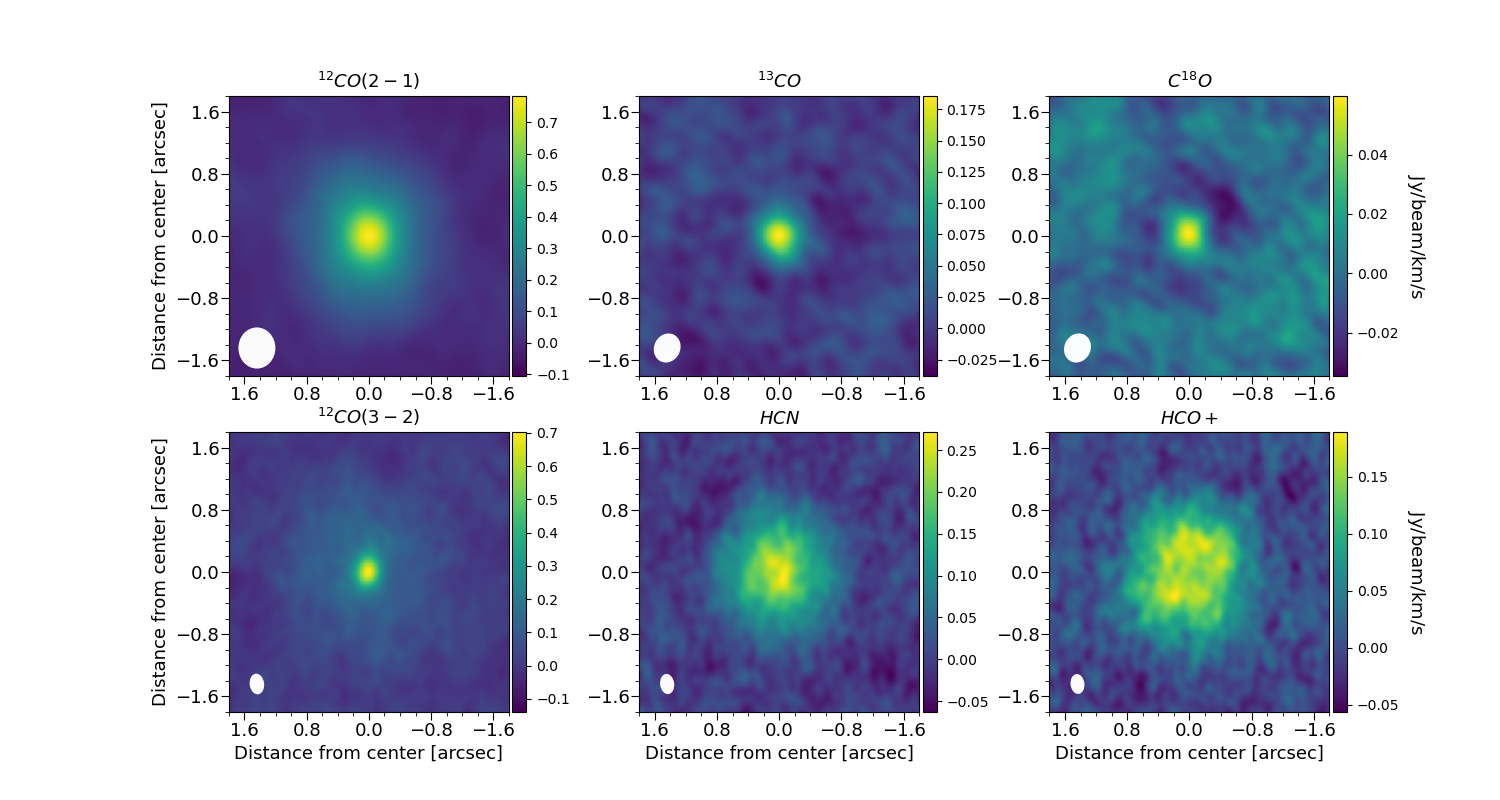}
\caption{Moment 0 Map of gas lines detected in Band 6 and Band 7 data.}
\label{fig:Image Lines B6-B7}
\end{figure*}

\FloatBarrier

\section{Line radial profiles}
\label{sec:C}

\begin{figure*}[h!]
\centering
\includegraphics[width=18cm]{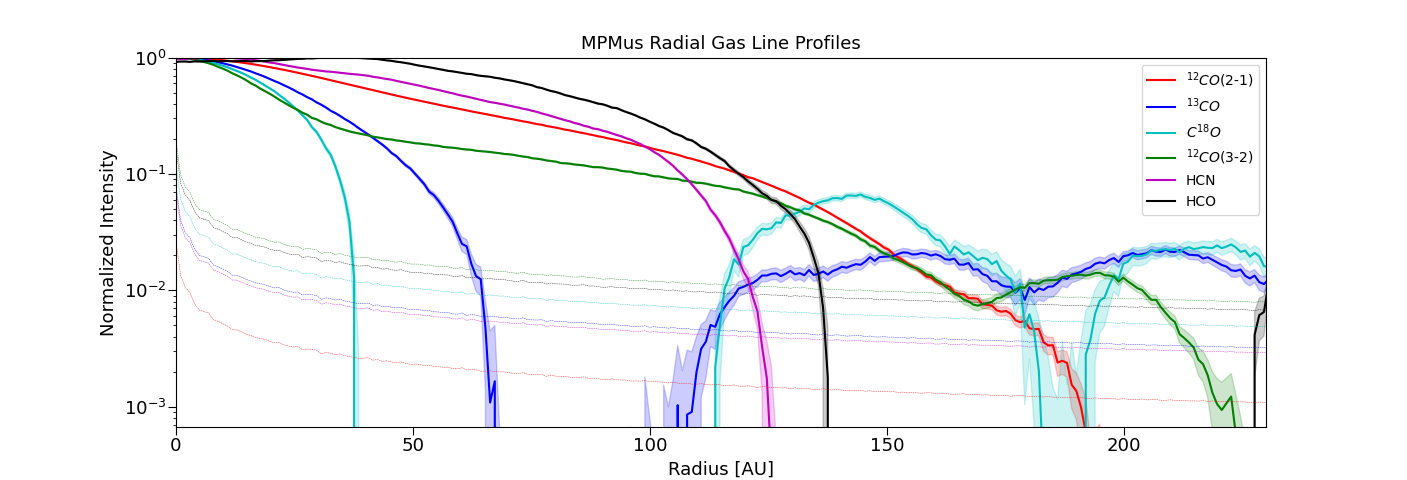}
\caption{Radial brightness profile of the continuum and emission/absorption lines for MP Mus, combining the extended and compact configurations in band 6.}
    \label{fig:Radial profile (ext+cont)}
\end{figure*}

\FloatBarrier
\twocolumn

\section{Gas disk radii}
\label{sec:D}

We measured the gas size by using the lines we had available. For this, we deprojected moment 0 images using position angle and inclination previously estimated by a 2D Gaussian fit of the continuum and then, extracted deprojected azimuthally averaged radial profiles of the gas emission lines. We considered the following definition for integrated disk flux:

\begin{equation}
    F(r) = 2\pi \int_{0}^{r} I(s)sds
\end{equation}

where $s$ is the projected radial coordinate in the sky and $I(s)$ is the observed intensity. We estimated the gas disk radius, $R_{gas}$, as the radius that contains 95$\%$ of the total flux. We obtained an $R_{^{12}CO(2-1)}$ of 110.8 $\pm$ 8.9 au and an $R_{^{12}CO(3-2)}$ of 121.7  $\pm$ 9.7 au.

\end{document}